\documentclass[a4paper,fleqn,usenatbib]{mnras}

\usepackage[T1]{fontenc}
\usepackage{ae,aecompl}

\usepackage{ulem}
\usepackage{graphicx}

\usepackage{txfonts}

\usepackage{hyperref}
\hypersetup{colorlinks=true, linkcolor=red, citecolor=blue, urlcolor=blue}

\usepackage{color}
\usepackage{amstext}
\usepackage{multirow}

\newcommand*\lephare{L\textsc{e} P\textsc{hare}}

\newcommand*\ylm{\mathit{lm}}
\newcommand*\om{\mathit{hM}}
\newcommand*\um{\mathit{UM}}

\newcommand*\NUVr{\mathrm{(NUV-r)}^{0}}
\newcommand*\rKs{\mathrm{(r-K_s)}^{0}}
\newcommand*\Ks{\mathrm{K_{s}}}

\title[Environment drives the fast quenching of  low-mass galaxies]{
On the fast quenching of young low-mass galaxies up to \textit{z} $\sim$ 0.6: new spotlight on the lead role of environment
}

\author[T. Moutard et al.]
{Thibaud Moutard,$^{1}$\thanks{tmoutard@ap.smu.ca}
Marcin Sawicki,$^{1}$\thanks{Canada Research Chair}
St\'ephane Arnouts,$^{2}$
Anneya Golob,$^{1}$ \newauthor 
Nicola Malavasi,$^{3}$
Christophe Adami,$^{2}$
Jean Coupon,$^{4}$
and Olivier Ilbert $^{2}$
\\
$^{1}$Department of Astronomy \& Physics and Institute for Computational Astrophysics, Saint Mary's University, 923 Robie Street, Halifax,  \\
~~Nova Scotia, B3H 3C3, Canada \\
$^{2}$Aix Marseille Universit\'e, CNRS, LAM - Laboratoire d'Astrophysique de Marseille, 38 rue F. Joliot-Curie, F-13388, Marseille, France \\
$^{3}$Department of Physics and Astronomy, Purdue University, 525 Northwestern Avenue, West Lafayette, IN 47907, USA \\
$^{4}$Astronomical Observatory of the University of Geneva, ch. d'Ecogia 16, 1290 Versoix, Switzerland 
}

\date{Accepted XXX. Received YYY; in original form ZZZ}

\pubyear{2017}

\begin{document}
\label{firstpage}
\pagerange{\pageref{firstpage}--\pageref{lastpage}}
\maketitle



\begin{abstract}
We investigate the connection between environment and the different quenching channels that galaxies are prone to follow in the rest-frame NUVrK colour diagram, as identified by \citet{Moutard2016b}.
Namely, the fast quenching channel followed by \textit{young} low-mass galaxies and the slow quenching channel followed by \textit{old} high-mass ones.
We make use of the >22 deg$^2$ covered the VIPERS Multi-Lambda Survey (VIPERS-MLS) to select a galaxy sample complete down to stellar masses of $M_* > 10^{9.4} M_\odot$ up to $z \sim 0.65$ ($M_* > 10^{8.8} M_\odot$ up to $z \sim 0.5$) and including 33,500 (43,000) quiescent galaxies properly selected at $0.2 < z < 0.65$, while being characterized by reliable photometric redshifts ($\sigma_{\delta z/(1+z)} \leq 0.04$) that we use to measure galaxy local densities. 
We find that (1) the quiescence of low-mass [$M_* \leq  10^{9.7}  M_\odot$] galaxies requires a strong increase of the local density, which confirms the lead role played by environment in their fast quenching and, therefore, confirms that the low-mass upturn observed in the stellar mass function of quiescent galaxies is due to \textit{environmental quenching}.
We also observe that (2) the reservoir of low-mass star-forming galaxies located in very dense regions (prone to environmental quenching) has grown between $z \sim 0.6$ and $ z \sim 0.4$ 
whilst the share of low-mass quiescent galaxies (expected to being environmentally quenched) may have simultaneously increased, 
which would plead for a rising importance of \textit{environmental quenching} with cosmic time, compared to \textit{mass quenching}.
We finally discuss the composite picture of such environmental quenching of low-mass galaxies and, in particular, how this picture may be consistent with a \textit{delayed-then-rapid} quenching scenario.
\end{abstract}


\begin{keywords}
galaxies: photometry -- 
galaxies: distances and redshifts -- 
galaxies: statistics --
galaxies: interactions -- 
galaxies: star formation --
galaxies: evolution 
\end{keywords}




\section{Introduction}
\label{introduction}

The fact that galaxies can be classified according to their star-formation activity into a blue/star-forming population, mostly made of disc galaxies, and a red/quiescent population, mainly consisting of elliptical galaxies, has been extensively documented in the last decade \citep[e.g.,][]{Hogg2003, Kauffmann2004, Baldry2006, Haines2007, Williams2009, Arnouts2013, Moutard2016a, Pacifici2016a}.
This bimodality, which can be observed to redshift $z \sim 4$ \citep[e.g.,][]{Ilbert2013, Muzzin2013, Tomczak2014, Mortlock2015, Davidzon2017}, is the statistical expression of a fairly rapid phenomenon of star-formation shutdown, the so-called \textit{quenching}.
The processes that are involved in such quenching of star formation are, however, still a matter of debate. In particular, the quenching mechanism(s) that turn(s) star formation off in low-mass galaxies may be quite different from what is at play in massive galaxies.

Now well established, the predominance of the quiescence in massive galaxies \citep[see, e.g.,][]{Bundy2006, Ilbert2010, Baldry2012, Davidzon2013, Moutard2016b} underlies a \textit{downsizing} of the star-formation quenching (i.e., the more massive a galaxy is, the earlier its star formation stops, on average). Furthermore, the high constancy of the stellar mass function (SMF) of star-forming galaxies at high mass supports the idea that star-formation activity is preferentially impeded above a given stellar mass \citep[i.e., the star-formation efficiency declines exponentially above this stellar mass;][]{Ilbert2010, Peng2010}, which has been confirmed to be remarkably stable over several Gyrs from $z \sim 1$ \citep[namely, $\mathcal{M}_\textsc{sf}^\star = 10^{10.64 \pm 0.01} M_{\odot}$ at $0.2 < z < 1.5$;][]{Moutard2016b}. Actually, this characteristic stellar mass may also be considered as a dark matter halo critical mass of $M_h \sim 10^{12} M_{\odot}$ \citep[assuming a stellar-to-halo mass ratio; e.g.,][]{Coupon2015}. This may be consistent with virial shock-heating processes \citep[e.g.,][]{Keres2005, DekelBirnboim2006, Cattaneo2006}, but other mechanisms able to halt the cold-gas supply such as feedback from a radio-loud active galactic nucleus (AGN) may also explain the star-formation quenching in massive galaxies \citep[e.g.,][]{Best2005, Croton2006, Karouzos2014}, which appears to be characterised by quite long timescales (of 1-to-a few Gyrs) over the last ten Gyrs \citep[e.g.,][]{Schawinski2014, Ilbert2015, Moutard2016b, Pandya2017}.
However, such \textit{mass quenching} processes can not be invoked in low-mass galaxies, and environmental effects have been put forth to explain the star-formation suppression in these galaxies.

Indeed, the latest measurements of the SMF reveal a clear excess of low-mass quiescent galaxies, which underlies an upturn around stellar masses of $M_* \sim 10^{9.7}  M_\odot$ observed in the local Universe \citep[e.g.,][]{Baldry2012, Moustakas2013} and at low redshift \citep[][to $z \sim 0.5$]{Drory2009, Moutard2016b}, and whose build-up is observed at higher redshift \citep[e.g.,][to $z \sim 1$ for the later]{Muzzin2013,Tomczak2014}. 
We recently showed that quiescent galaxies that are responsible for this low-mass upturn in the SMF are \textit{young} quiescent galaxies \citep{Moutard2016b} --i.e. they exhibit colours typical of young stellar populations  \citep[making them good candidates to being post-starburst galaxies; e.g.,][]{Kriek2010, Whitaker2012}-- and are expected to have experienced a rapid quenching (turning quiescent over just $\sim 0.4$ Gyr). 
Such observations support a picture where galaxies follow different quenching channels depending on their stellar mass, which is consistent with a scenario mixing different modes of star-formation quenching as proposed by \citet{Faber2007}.
In particular, the excess of low-mass quiescent galaxies has been suggested to be associated with the \textit{environmental quenching} of satellite galaxies, whose importance is expected to grow with large-scale structure and, thus, to decrease with increasing redshift \citep{Peng2010}. 

The connection between environment and star-formation quenching is now well illustrated at low redshift \citep[$z < 0.5$; e.g.,][]{Balogh1997, Lewis2002, Hogg2003, Kauffmann2004, Baldry2006, Haines2007, Yang2009, Peng2012}, and a clear picture has emerged where, on average, red/quiescent galaxies are characterised by richer/denser environments than blue/star-forming ones.
Several quenching processes involving rich environments have therefore been proposed, such as \textit{ram-pressure stripping}, in which the gas is expelled from the galaxy that becomes satellite \citep[][]{Gunn1972}; \textit{strangulation/starvation},\footnote{We emphasise that the terms \textit{strangulation/starvation} might either refer to environment (e.g., when a galaxy enters the hot gas of a cluster) or to peculiar evolution (e.g., when the radio-loud AGN feedback halts the cold gas infall).} 
in which the cold gas supply is heated and then halted \citep[][]{Larson1980, Peng2015}; galaxy harassment, in which multiple encounters deprive galaxies from stars and/or gas through tidal stripping \citep[][]{Farouki1981, Moore1996}; or \textit{major merging} triggering a subsequent starburst episode and/or an AGN that consumes/expels the remaining reservoir of cold gas \citep[][]{Schawinski2014}; all assuming that cold gas fuelling is impeded in dense environments. 
This stresses indeed that these processes must be addressed in the cosmological context of the hierarchical growth of large-scale structures, and especially the evolution of filaments along which flows the cold gas that fuels star formation inside galaxies \citep{Sancisi2008, Dekel2009}.

Much effort has been made over the last decade to observe the impact of environment on star formation across cosmic time \citep[e.g.,][]{Cucciati2006, Cucciati2010, Muzzin2012, Lani2013, Scoville2013, Muzzin2014, Fossati2017, Cucciati2017, Malavasi2017, Laigle2018}. However, for different reasons, these studies focused on relatively massive galaxies and were not able to probe a low-mass population whose prime interest is precisely the fact that, by "nature", it is not expected to quench. 
While the impact of environment on the quenching of low-mass quiescent galaxies has been observed for a while in the local Universe \citep{Hogg2003} where these galaxies have appeared to be essentially satellites \citep[e.g.,][]{Haines2007, Peng2012}, the impact of environment on the quenching of low-mass galaxies was not observed at higher redshift until recently \citep[namely, at $0.5 < z < 1$; ][]{Guo2017}.
This reemphasized the question of the impact of environment on the quenching of low-mass galaxies across cosmic time while raising the question of the associated contribution to the build-up of the quiescent population, in particular, in the light of the picture described previously where galaxies are prone to follow different quenching channels depending on their stellar mass.

In this paper, we analysed the relation between environment and the different quenching channels that galaxies are prone to follow in the rest-frame NUV-r vs. r-K colour diagram, notably depending on their stellar mass.
In particular, we intended to verify whether environment drives the fast quenching channel followed by low-mass galaxies and responsible for the upturn observed in the SMF of quiescent galaxies, as shown in \citet[][]{Moutard2016b}. At the same time, we took this opportunity to question the  importance of such quenching channel across cosmic time, compared to the quenching channel that can be associated with mass quenching.
We made use of the unique combination of area, depth and photometric multi-wavelength coverage of the VIPERS Multi-Lambda Survey\footnote{\texttt{http://cesam.lam.fr/vipers-mls/}} \citep[VIPERS-MLS;][]{Moutard2016a}, assembled in the fields of the VIMOS Public Extragalactic Redshift Survey\footnote{\texttt{http://vipers.inaf.it/}} \citep[VIPERS;][]{Guzzo2014}. Covering >22 deg$^2$ down to $\Ks < 22$, the VIPERS-MLS is indeed remarkable as
(\textit{a}) it allows the use of the rest-frame NUV-r vs. r-K (NUVrK) diagram to properly separate quiescent and star-forming galaxies;
(\textit{b}) it provides a complete sample of galaxies down to stellar masses of $M_* \simeq 10^{9.4} M_{\odot}$ at $z < 0.65$ ($M_* \simeq 10^{8.8} M_{\odot}$ at $z < 0.5$) including more than 33,500 (43,000) quiescent galaxies, which enabled us to probe the evolution of fairly low-mass galaxies from $z \sim 0.6$;
while (\textit{c}) these galaxies are all characterised by accurate photometric redshifts, with $\sigma_{\delta z/(1+z)} < 0.04$, which allows for reliable local density measurements.

The paper is organised as follows. In Sect. \ref{sect_VMLS} we give an overview of the VIPERS-MLS data and measurements used in the present study. We then review the NUVrK diagram and its ability to distinguish between fast and slow quenching channels in Sect. \ref{sect_NUVrK}. In Sect. \ref{sect_results} we present our results regarding the connection between environment and quenching channels to finally discuss these results in Sect. \ref{sect_discuss}.

Throughout this paper, we use the standard cosmology ($\Omega_m~=~0.3$, $\Omega_\Lambda~=~0.7$ with $H_{\rm0}~=~70$~km~s$^{-1}$~Mpc$^{-1}$). Magnitudes are given in the $AB$ system \citep{Oke1974} and galaxy stellar masses are given in units of solar masses ($M_{\odot}$) for a \citet{Chabrier2003} initial mass function.

\section{Data: VIPERS-MLS}
\label{sect_VMLS}

Observational data, photometric redshifts and stellar mass estimates were discussed extensively in \citet{Moutard2016a, Moutard2016b} and here we only present a brief overview of key elements (Sect. \ref{sect_data}).  
To these preexisting measurements, we have now also added the measurement of local galaxy density, as described in Sect. \ref{sect_dens_measur}. 

\subsection{Observational data, photometric redshifts, and mass estimates}
\label{sect_data}

Our data consist of (FUV, NUV,) u, g, r, i, z and $\Ks$ imaging of 22.38 deg$^2$ --after masking and quality cuts-- within the VIPERS-MLS \citep{Moutard2016a}, a follow-up program in the fields of the spectroscopic survey VIPERS \citep{Guzzo2014}, i.e., in the fields W1 and W4 of the Canada-France-Hawaii Telescope Legacy Survey\footnote{\texttt{http://www.cfht.hawaii.edu/Science/CFHTLS/}} (CFHTLS).
The VIPERS-MLS optical imaging has been based on the CFHTLS T0007 release \citep{Hudelot2012} that reaches 80per cent completeness depth to i$\sim 23.7$, while the $\Ks$-band data were obtained through new observations reaching $\Ks \sim 22$ over $\sim27$ deg$^2$. 
The multi-wavelength coverage of the VIPERS-MLS has been complemented by GALEX \citep{Martin2005} FUV and NUV data combining preexisting and new observations over $\sim12.7$ deg$^2$, incorporated after using u-band images as priors. 
For full details of the data processing and catalogue creation see \citet{Moutard2016a}. 

Photometric redshifts were derived as described in \citet{Moutard2016a} by using the template-fitting code \lephare \ \citep{Arnouts2002, Ilbert2006}. Photometric redshift (photo-z) estimates were validated using extensive VIPERS spectroscopy \citep[$\sim$90,000 spectroscopic redshifts to i $=22.5$;][]{Scodeggio2018}, combined with smaller numbers of high-quality redshifts taken from deeper spectroscopic datasets. Their accuracy is characterized by $\sigma_{\delta z/(1+z)} \sim 0.03$ to i $<22.5$ and $\sigma_{\delta z/(1+z)} \sim 0.05$ for i $>22.5$ galaxies, with corresponding catastrophic outlier rates of $\eta=1.2$ per cent and $\eta=9$ per cent \citep[][Figure 3]{Moutard2016b}. In the case of the faintest galaxies we considered in this paper, namely low-mass quiescent galaxies with $M_* < 10^{9.7}  M_\odot$ around $z \sim 0.65$, the photo-z accuracy is better than $\sigma_{\delta z/(1+z)} \sim 0.04$. Star/galaxy separation \citep[described in][]{Moutard2016a} discarded 97 per cent of stars while keeping 99 per cent of galaxies. 

Galaxy stellar masses were derived as described in \citet{Moutard2016b} with \lephare \ using dust-corrected \citet[][hereafter BC03]{BC2003} models of spectral energy distribution (SED), modified to include the effects of emission lines. Rest-frame colours were computed using the nearest observed-frame band in order to minimize dependence on model spectra.
The depth of our data allows us to push our analysis to galaxies with $M_* > 10^{8.8} M_\odot$ around $z \sim 0.5$, and $M_* > 10^{9.4} M_\odot$ around $z \sim 0.65$.

\subsection{Measurement of the local density}
\label{sect_dens_measur}

\begin{figure}

\includegraphics[width=\hsize, trim = 0cm 0cm 0cm 0cm, clip]{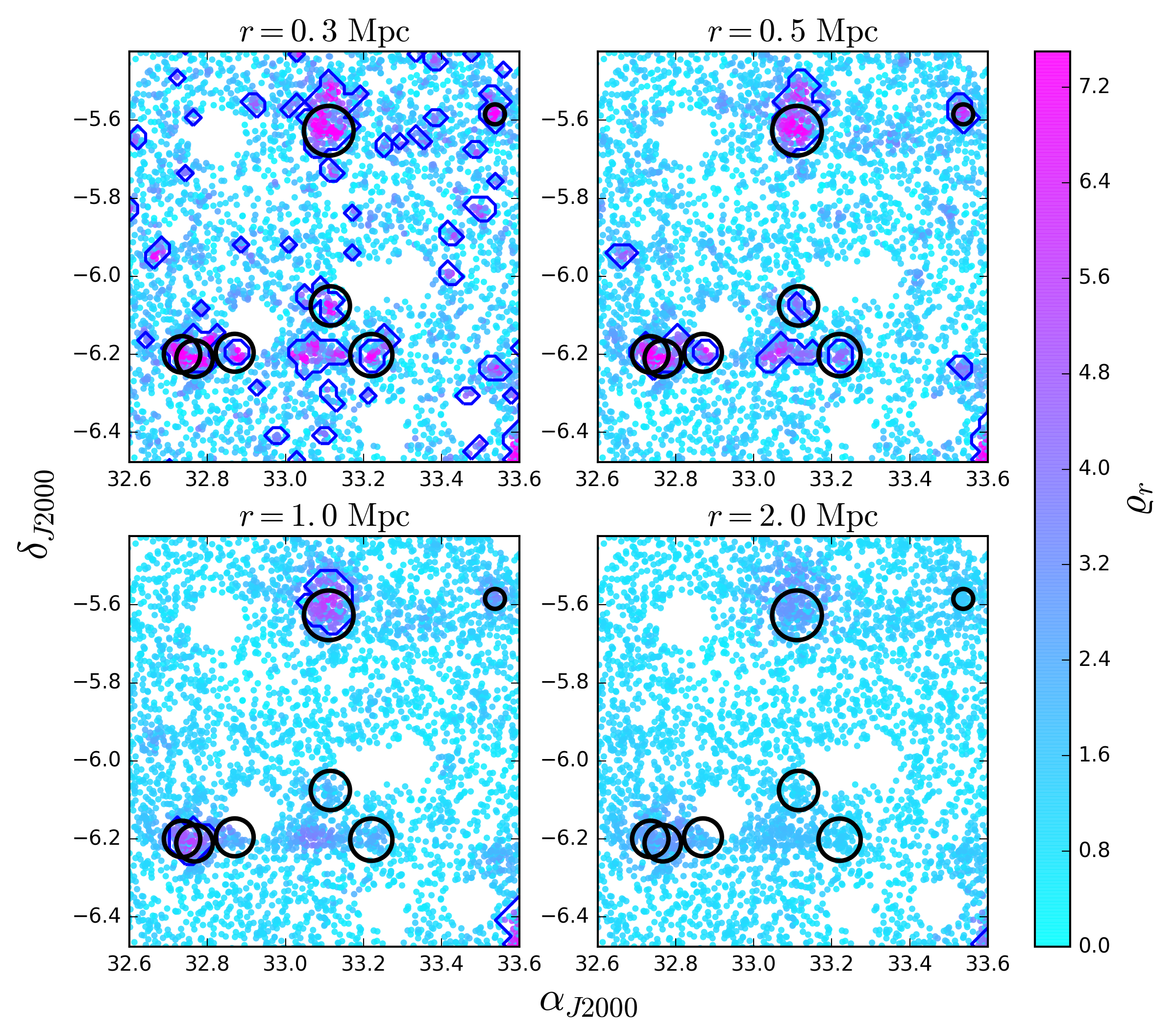}

\caption{Map of the local density $\varrho_r$ at $0.2 < z < 0.5$, as measured in one $1\times 1$ deg$^2$ patch of the VIPERS-MLS for different aperture radii $r$ of 0.3, 0.5, 1 and 2 physical Mpc. Each point indicates the position of one galaxy and the colour codes the corresponding local density, as derived from Eq. \ref{eq_loc_dens}, while blue contours delineate regions where $\varrho_r > 4$. Black circles reflect the position and size of bright X-ray clusters identified in the XXL survey \citep{Pacaud2016}.
\label{fig_dens_aper}}
\end{figure}

To measure the local density of environment surrounding galaxies in our photo-z sample, we adopted a method similar to \citet{Lani2013} and \citet{Malavasi2016}. In brief, we counted the number of galaxies lying in a cylinder of fixed aperture centered on each galaxy for which we measured the density. The cylinder physical depth was set at 1 Gyr, which turned out to be a good compromise to avoid galaxy exclusion and excessive dilution in our case\footnote{A depth of 1 Gyr represents a redshift depth of $\Delta z \simeq 0.13-0.15$ around $z \sim 0.65$, i.e., $\sim 2 \ \times$ the typical photo-z uncertainty affecting the faintest (quiescent) galaxies of our sample (namely, $\sigma_{\delta z/(1+z)} \simeq 0.04$) 
and a redshift depth of $\Delta z \simeq 0.1-0.13$ at $z \lesssim 0.5$ (where $ \sigma_{\delta z/(1+z)} \simeq 0.03$), which then corresponds to $\sim3 \  \times$ the photo-z uncertainty of our faintest galaxies.} \citep[for a detailed analysis of the completeness and purity associated with the use of photometric redshifts for reconstructing the galaxy density field, please refer to][]{Malavasi2016}.

It is convenient to define $\varrho$, as the local density $\rho$ normalized by the mean density of the Universe $\langle \ \rho \  \rangle$ at the same redshift:
\begin{equation}
\varrho = \frac{\rho}{\langle \ \rho \  \rangle} = 1 + \delta
\end{equation}
where one can see that $\varrho$ can also be expressed in terms of the density contrast $\delta$, as it is commonly defined.\footnote{The density contrast is defined by $\delta = \frac{\rho \ - \ \langle \ \rho \  \rangle}{\langle \ \rho \  \rangle}$.}
Quoting $\varrho_r$ the normalized local density measured in cylinder of aperture radius $r$, we can write
\begin{equation}
\label{eq_loc_dens}
\varrho_r  = \frac{n_r \ / \ a_r}{N \ / \ A} \ ,
\end{equation}
where $n_r$ is the number of surrounding galaxies within a cylinder of aperture radius $r$ and effective (i.e., non masked) area $a_r$, and
$N$ is the total number of galaxies within the corresponding 1 Gyr redshift layer over the entire effective area of the survey $A$ (namely, $22.38$ deg$^2$ in the present analysis\footnote{We used the  $\Ks$ part of the VIPERS-MLS.}).
We emphasize that, while we made use the normalized local density $\varrho$ in the present study, it is generally simply referred as the "local density" in the following, for sake of simplicity.

\begin{figure*}
\center
\includegraphics[width=0.95\hsize, trim = 0.0cm 0cm 0.0cm 0cm, clip]{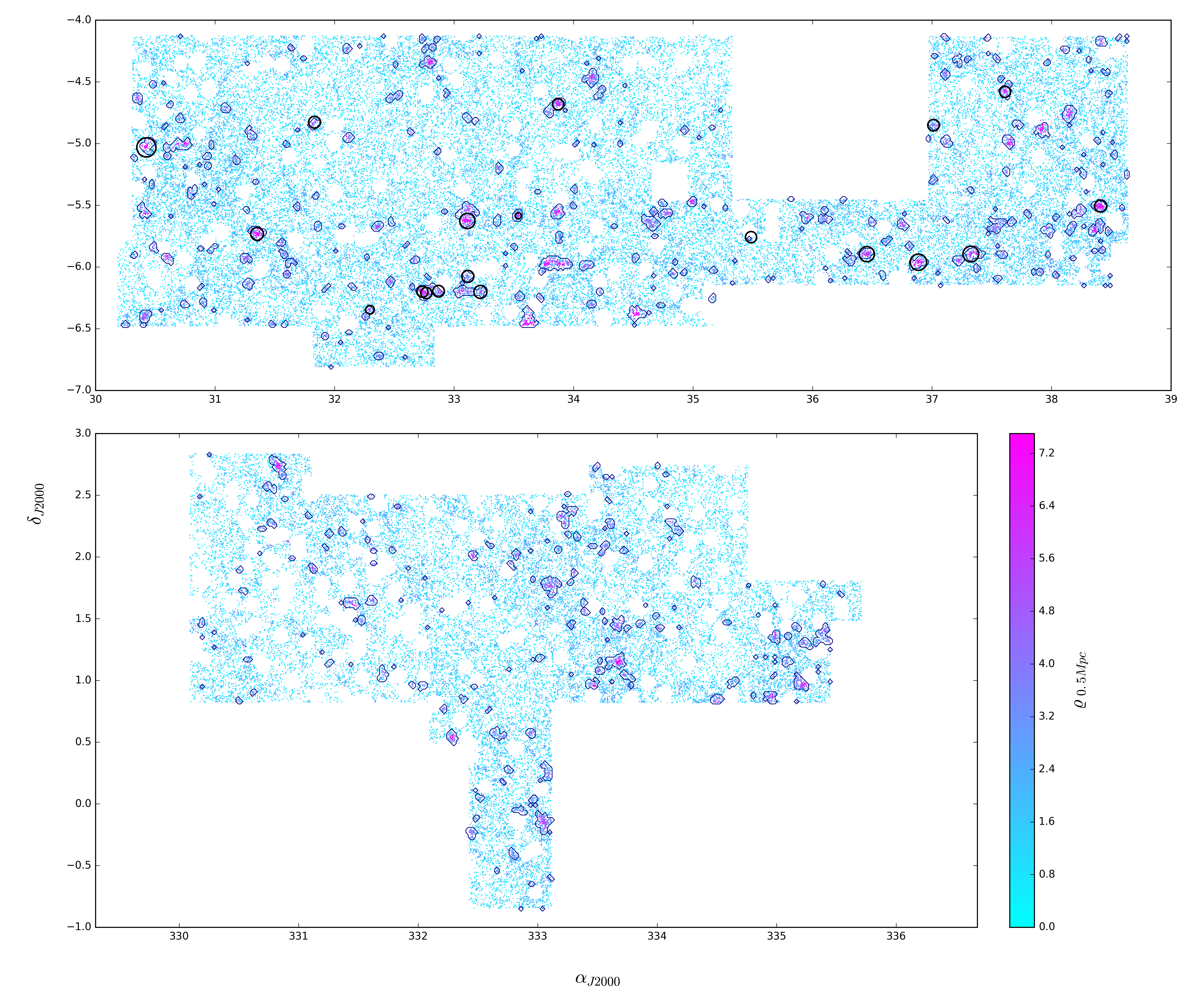}
\caption{Map of the local density measured at $0.2 < z < 0.5$ in the fields W1 (top) and W4 (bottom) of the VIPERS-MLS. As in Fig. \ref{fig_dens_aper}, each point shows the position of a galaxy while the colour codes $\varrho_\mathrm{\ 0.5 \ Mpc}$, the local density measured in 0.5 Mpc radius apertures around each galaxy, and dark-blue contours outline regions where $\varrho_\mathrm{\ 0.5 \ Mpc} > 4$ (associated with massive optical clusters). 
Black open circles shows the position of bright X-ray clusters identified as part of the XXL survey, for comparison (only available in W1).
\label{fig_dens_footprint}}
\end{figure*}

Aiming to better take advantage of the angular information, we tried several cylinder apertures with radii ranging from 0.3 to 2 physical Mpc (i.e., around the typical galaxy cluster size in the considered redshift range). 
Figure \ref{fig_dens_aper} shows the local density measured in one $1\times 1$ deg$^2$ patch of the VIPERS-MLS where we can compare with a map of relaxed galaxy clusters from the XXL survey \citep{Pierre2016}. 
Namely, we made use of $\sim 100$ confirmed X-ray clusters from the XXL bright cluster sample \citep{Pacaud2016}. Selected with a flux lower limit of $3 \times 10^{-14} erg s^{-1} cm^{-2}$ in the $[0.5-2]$ keV band of the XMM-Newton satellite, most of XXL bright clusters have masses $7 \times 10^{13} M_\odot \leq M_{500} \leq 3 \times 10^{14} M_\odot$ and redshifts $0.1 \leq z \leq 0.5$. 
Unsurprisingly, as one can see, large apertures ($r \geq 1$ Mpc) tend to smooth the density field, while small apertures ($r \lesssim 0.3$ Mpc) may provide noisier measurements of $\varrho_r$. 
We verified that a radius of 0.5 Mpc appears to be a good compromise enabling the detection of over-dense regions whose angular distribution and size match that of bright X-ray clusters, while preventing the measured local density field from being too noisy\footnote{This may typically happen when a few galaxies are concentrated over tiny effective areas.} by ensuring that over-densities are basically defined from a significant number of galaxies (typically, $\gtrsim 20$ galaxies for r$=$0.5 Mpc).
As a matter of fact, by considering over-dense regions where $\varrho_\mathrm{\ 0.5 \ Mpc} > 4$ (blue contours in Fig. \ref{fig_dens_aper}), we recover 19/19 XXL bright clusters lying in the VIPERS-MLS field at $0.2 < z < 0.5$ (and 2/2 at $0.5 < z < 0.65$).\footnote{Matching was performed according to both the angular position and the redshift, within an angular radius corresponding to XXL cluster $M_{500}$ radii and by considering the median angular position of our detected over-dense regions, while the redshift tolerance was determined by the typical photo-z uncertainties associated with the median redshift of the over-dense regions.}


Figure \ref{fig_dens_footprint} shows the local density map in the two fields of the VIPERS-MLS at $0.2 < z < 0.5$,
as measured in 0.5 Mpc radius apertures. The use of photo-z prevents us from being able to trace the subtleties of large-scale structures, like filaments. On the other hand, the method enables the detection of the most massive structures such as clusters (typically seen with $\varrho_{\ 0.5 \mathrm{Mpc}} > 4$, as shown in Fig. \ref{fig_dens_aper}).
As for over-densities having no bright XXL counterparts where the VIPERS-MLS and XXL survey overlap, we cannot exclude some of them to be artefacts. For example, due to the fact that our measure of the local density is projected along the cylinder depth, some large-scale structures may organise along the line of sight (e.g., filament or pair of overlapping groups/clusters).

However, the contribution of such alignments is expected to be low in a large-scale survey
and, as shown and discussed in previous studies \citep[see, e.g.,][]{Muldrew2012, Haas2012, Lani2013, Malavasi2016}, the uncertainties associated with the use of photometric redshifts tend, on the contrary, to dilute real over-densities along the line of sight,\footnote{In the absence of any significant photometric-redshift bias or catastrophic outlier rate, which is the case in the present study.} which therefore makes fake detection of over-dense regions even less probable in our analysis.
Moreover, some of these clusters may be not yet virialised clusters (i.e., they are faint or not X-ray emitters) or, even, simply not part of this early XXL release.

We verified that our results were self-consistent by measuring local densities using an alternative approach based on distances to the $n^{th}$-nearest neighbour (typically when $n = 7$).
At the same time, densities based on fixed aperture deal naturally well with masked areas (critical in W4) and were shown to correlate very well with high-mass halos \citep[more precisely, when the aperture diameter scales with the virial radius of the halo, typically, $<1$ Mpc;][]{Muldrew2012, Haas2012}, which is well suited to our analysis (where the use of photometric redshifts 
allows the detection of fairly massive galaxy clusters).

\section{The NUVrK diagram as a tracer of galaxy evolution}
\label{sect_NUVrK}


As shown by \cite{Arnouts2013}, the rest-frame NUV--r vs. r--K diagram (hereafter NUVrK diagram) is a powerful alternative to the rest-frame UVJ diagram \citep{Williams2009} to separate quiescent (Q) galaxies from (very dusty) star-forming (SF) galaxies.
By extending the wavelength scope of the SED from NUV to NIR, the NUVrK diagram is indeed more sensitive to instantaneous SFR\footnote{UV emission is sensitive to the lifetime of B/A stars, i.e., $10^{-2}$--$10^{-1}$ Gyr.} while being sensitive to stellar ageing and dust attenuation. This results in the enlargement of the so-called \textit{green valley}, i.e, the space that separates SF and Q galaxies, which allows for a robust selection of star-forming, quiescent, and transitioning (i.e., quenching) galaxies.

\subsection{Identifying two quenching channels in the NUVrK diagram}

\begin{figure*}
\center
\includegraphics[width=0.99\hsize]{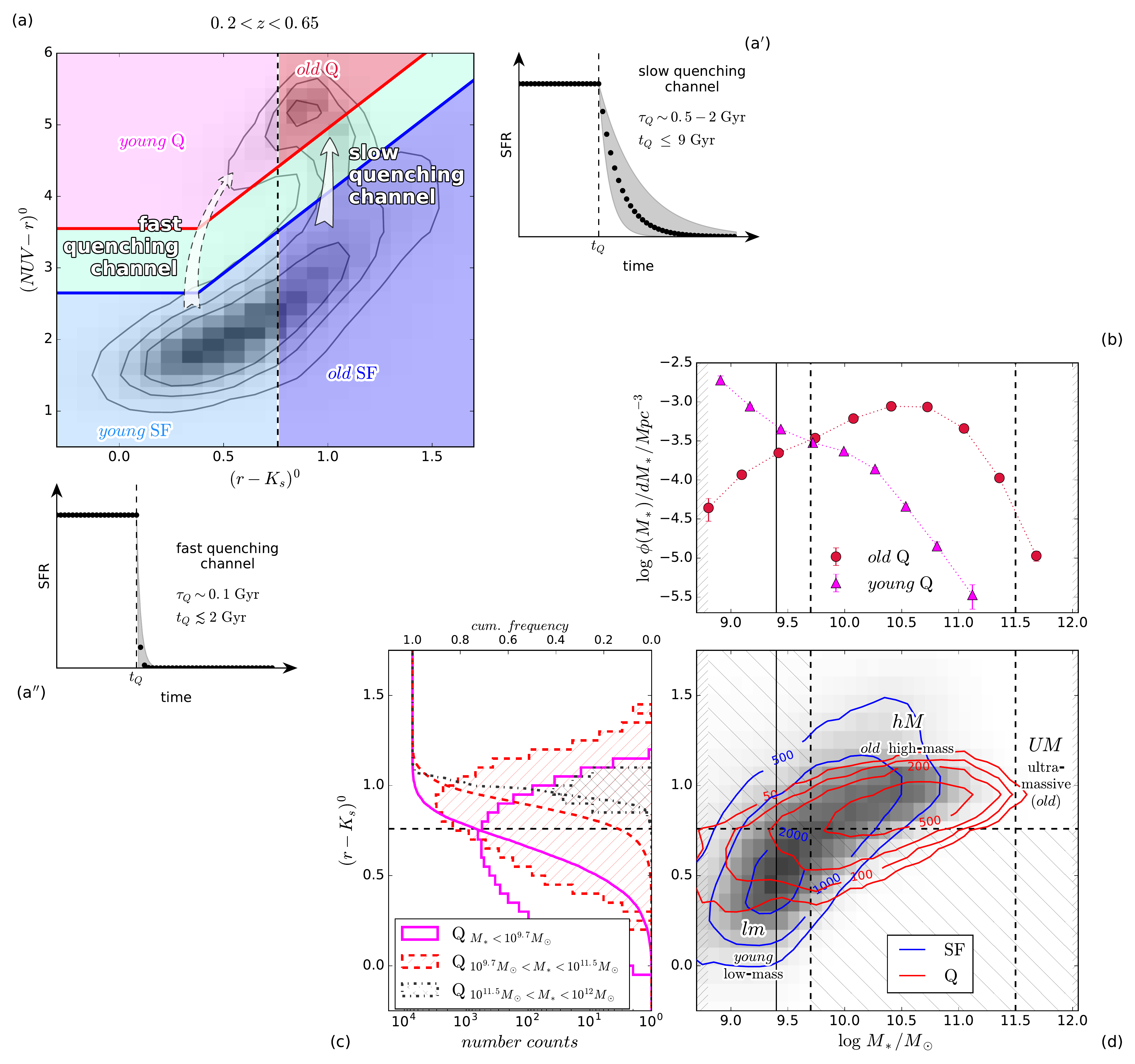}

\caption{NUVrK/stellar-mass selection scheme of the different classes of galaxies adopted in our analysis.
\textbf{(a)} Galaxy distribution at $0.2 < z < 0.65$ in the NUVrK diagram, which allows for the selection of quiescent (Q) galaxies (above the red line; see Eq. \ref{eq_Qsel}) and star-forming (SF) ones (below the blue line; see Eq. \ref{eq_SFsel}) on both sides of the so-called \textit{green valley} (in green).
The rest-frame colour cut at $\rKs = 0.76$ (vertical black dashed line) enables the separation between \textit{old} galaxies (right), prone to slow quenching, and \textit{young} galaxies (left), susceptible to fast quenching. 
Corresponding models of star-formation history (SFH) are shown in subpanels a$^{\prime}$ and a$^{\prime\prime}$, respectively:
namely, constant star-formation rate (SFR) until the time of the quenching $t_Q$, followed by an exponential decline with form SFR$(time) \varpropto e^{-(time-t_Q)/\tau_Q}$, where the slow quenching of \textit{old} galaxies is characterized by fairly long time-scales of $\tau_Q \sim 0.5-2$ Gyrs while \textit{young} galaxies are characterized by quenching time-scales of $\tau_Q \sim 0.1$ Gyr \citep{Moutard2016b}. 
\textbf{(b)} Stellar mass function (SMF) of \textit{old} (red circles) and \textit{young} (magenta triangles) quiescent galaxies, as defined in panel a.
\textbf{(c)} Number counts and cumulative frequencies in $\rKs$ for quiescent galaxies of the stellar-mass bins described in Sect. \ref{sec_sample_sel}: [$M_* < 10^{9.7} M_\odot$] (magenta solid lines), [$10^{9.7} < M_*/M_\odot < 10^{11.5}$] (red dashed lines) and [$10^{11.5} < M_*/M_\odot < 10^{12}$] (grey dot-dashed lines).
\textbf{(d)} Galaxy distribution at $0.2 < z < 0.65$ in the stellar-mass vs. $\rKs$ plan, and corresponding selection of \textit{young} low-mass galaxies ($\ylm$, Eq. \ref{eq_ylm}; responsible for the upturn observed in the quiescent SMF and susceptible to fast quenching), \textit{old} high-mass galaxies ($\om$, Eq. \ref{eq_om}; responsible for the quiescent SMF build-up around $\mathcal{M}^\star _{\textsc{sf}} \sim 10^{10.64} M_\odot$ and prone to quench slowly) and ultra-massive galaxies ($\um$, Eq. \ref{eq_um}; predominantly \textit{old} and quiescent at these redshifts). 
Isodensity contours for the star-forming population are reported in blue (namely for 500, 1000 and 2000 galaxies/pixel) and in red for the quiescent population (for 50, 100, 200, and 500 galaxies/pixel).
The vertical thin black solid line shows the stellar mass completeness limit considered to $z \sim 0.65$, with log $M_{lim}/M_\odot = 9.4$.
Hatched regions indicate the stellar-mass and rest-frame color regimes that are excluded from our study (see Sect. \ref{sec_sample_sel}). 
\label{fig_age_mass_sel}  }
\end{figure*}


As shown in \citet{Moutard2016a}, the volume probed in the VIPERS-MLS is well suited to probe rare populations, which may notably enable us to catch transitioning galaxies that were not observed in smaller surveys. This led us to identify a quenching channel followed by fairly massive galaxies \citep[typically when reaching stellar masses around the characteristic mass $\mathcal{M}^\star _{\textsc{sf}} \simeq 10^{10.64} M_\odot$;][]{Moutard2016b}, as recalled in the following.  

By quenching channel, we mean a pathway in the rest-frame NUVrK colour diagram that quenching galaxies follow from the star-forming population to the quiescent population.
A quenching channel can be associated with an average star-formation history (SFH), which may be highlighted through comparison between colour evolution tracks predicted by stellar-population synthesis models and the actual distribution of galaxy rest-frame colours \citep[see, e.g.,][]{Schawinski2014, Marchesini2014, Moutard2016b, Pacifici2016a}. 
In particular, the NUVrK diagram turns out to be very well suited to distinguish SFHs characterized by different quenching time-scales \citep[][Fig. 20]{Moutard2016b} for it is sensitive to very different star lifetimes on each of its axis:  $< 0.1$ Gyr along the rest-frame NUV--r colour \citep{Salim2005, Martin2007}, hereafter quoted $\NUVr$, and $> 1$ Gyrs along the rest-frame r--K colour \citep{Arnouts2007,Williams2009}, hereafter quoted $\rKs$. 
Indeed, $\NUVr$ traces recent star-formation (thanks to rest-frame NUV), while $\rKs$ results from the combination of stellar ageing (i.e., the accumulation of generations of low-mass stars, notably traced by rest-frame r) and dust extinction \citep[rest-frame r--$\Ks$ being a good tracer of the infrared excess, i.e., the ratio between the UV light absorbed by dust and its re-emission in the infrared;][]{Arnouts2013}: galaxy $\rKs$ colours are therefore expected to redden with cosmic time, on average.
Thus, in the NUVrK diagram, a galaxy experiencing an early and rapid quenching of the star formation will see its $\NUVr$ colour rapidly reddened (typically by $\sim1$ mag) whilst its $\rKs$ colour will simultaneously remain blue, while a slow quenching will be characterised by the slow reddening of both the $\NUVr$ and $\rKs$ colours.

In Fig. \ref{fig_age_mass_sel}a, we show the NUVrK distribution of our galaxy sample at $0.2 < z < 0.65$ and the corresponding selection of quiescent and star-forming galaxies, as defined on both sides of the so-called "green valley" where one can identify a line of transitioning (i.e., quenching) galaxies concentrated at $0.76 < \rKs< 1.23$ that turn out to be fairly massive ($>60$ per cent of galaxies with $10^{10.5} < M_* / M_\odot < 10^{11}$).
We used the upper and lower limits of the time-dependant selection of Q and SF galaxies defined in \citet{Moutard2016b}, so that galaxies in transition in the \textit{green valley} are excluded from our analysis.
Namely, Q galaxies were selected with
\begin{eqnarray}
\left[ ~\NUVr > 3.772 - 0.029 \times t_{\textsc{l}} ~\right] ~ \cap \nonumber\\
\left[ ~\NUVr  > 2.25 \times \rKs  + 2.768 - 0.029 \times t_\textsc{l} ~\right] 
\label{eq_Qsel}
\end{eqnarray}
and SF galaxies with 
\begin{eqnarray}
\left[ ~\NUVr < 2.922 - 0.029 \times t_{\textsc{l}} ~\right] ~ \cup \nonumber\\
\left[ ~\NUVr  < 2.25 \times \rKs  + 1.918 - 0.029 \times t_\textsc{l} ~\right]  \ ,
\label{eq_SFsel}
\end{eqnarray}
where $t_{\textsc{l}}$ is the look-back time at gigen redshift\footnote{E.g., $t_{\textsc{l}} \sim 4$ Gyrs at $0.2 < z < 0.5$ and $\sim 5.5$ Gyrs at $0.5 < z < 0.65$.} \citep[for more detail, please refer to][Sect. 5.1]{Moutard2016b}.

One can see how a conservative cut at $\rKs > 0.76$ allows us to isolate a population of \textit{old} quiescent galaxies --i.e., galaxies that exhibit colours typical of evolved (old and dusty) stellar populations-- that is expected to be fed by the quenching of fairly high-mass star-forming galaxies reaching $\sim \mathcal{M}^\star _{\textsc{sf}} = 10^{10.64} M_\odot$ (red points in Fig. \ref{fig_age_mass_sel}b), while quiescent galaxies exhibiting bluer $\rKs < 0.76$ colours --typical of younger stellar populations-- must have followed another quenching channel to turn quiescent (since galaxy $\rKs$ colours only redden with time, as explained above). 
Furthermore, as pointed out in \citet{Moutard2016b}, these \textit{young} quiescent galaxies are essentially low-mass galaxies, and are thus responsible for the upturn observed in the SMF of quiescent galaxies at $M_* < 10^{9.7}  M_\odot$ (magenta triangles in Fig. \ref{fig_age_mass_sel}b).


Actually, considering simple e-folding SFH models turns out to be qualitatively and quantitatively well adapted to approximate the average SFHs associated with the flux of quenching galaxies that cross the green valley \citep[see, e.g.,][]{Schawinski2014,Moutard2016b,Pacifici2016b}, the direction of this flux  being supported by the rising fraction of quiescent galaxies that observed with cosmic time from $z \sim 4$ \citep[e.g.,][]{Muzzin2013, Mortlock2015}.
Doing so, rejuvenation episodes\footnote{As observed in the local Universe \citep{Salim2010,Thomas2010} or predicted at higher redshift \citep{Trayford2016}.} that may affect the SFR --and so, the $\NUVr$ colour\footnote{SFR variations only affect $\NUVr$, which traces recent star-formation, not $\rKs$.}-- of individual galaxies are neglected.
Indeed, considering an average SFH is equivalent to stacking the SFHs of individual galaxies, which tends to smooth the stochasticity that may characterise their SFRs across cosmic time.
As detailed in \citet{Moutard2016b}, the average SFH models considered to explain the different quenching channels identified in the NUVrK diagram are characterised by a constant SFR while galaxies are on the star-formation main sequence, until the time of the quenching $t_Q$ where the SFR drops as SFR$(time) \varpropto e^{-(time-t_Q)/\tau_Q}$ with a quenching time-scales $\tau_Q$ (as represented in panels a$^{\prime}$ and a$^{\prime\prime}$ of Fig. \ref{fig_age_mass_sel}).

The quenching channel followed by \textit{old} high-mass galaxies has thus been shown to be characterised by fairly long time-scales of $\tau_Q \sim 0.5-2$ Gyrs (Figs. \ref{fig_age_mass_sel}a, \ref{fig_age_mass_sel}a$^{\prime}$).  
This corresponds to quenching durations of $\sim 1-3.5$ Gyrs to cross the green valley after leaving the main sequence,\footnote{The set of tracks presented in \citet[][Fig. 20]{Moutard2016b} is limited to the cases that are able to explain the presence of transitioning high-mass galaxies in the green valley, found to be concentrated at $0.76 < \rKs< 1.23$, which is also visible here in Fig. \ref{fig_age_mass_sel}a.} which is compatible with a \textit{strangulation} scenario where the cold gas inflows are impeded and the galaxy consumes slowly its reservoir of remaining gas \citep[see, e.g.,][]{Peng2015}. 
In contrast, the quenching channel that would allow us to explain the presence of \textit{young} low-mass quiescent galaxies must be characterised by shorter time-scales of $\tau_Q \sim 0.1$ Gyrs (see Figs. \ref{fig_age_mass_sel}a, \ref{fig_age_mass_sel}a$^{\prime\prime}$), which corresponds to quenching durations of only $\sim 0.4$ Gyrs to cross the green valley; as a corollary, \textit{young} low-mass quiescent galaxies are also characterised by a recent quenching.

The NUVrK diagram is therefore a powerful tool to explore the different quenching channels that may be followed by galaxies across cosmic time: fast for (\textit{young}) low-mass galaxies, slow for (\textit{old}) high-mass ones.

\subsection{Selection of (young) low-mass and (old) massive and ultra-massive galaxies}
\label{sec_sample_sel}

While \textit{young} [$\rKs$ < 0.76] quiescent galaxies are essentially low-mass galaxies, their stellar-mass distribution stretches to $\gtrsim 10^{10.5} M_\odot$ (Fig. \ref{fig_age_mass_sel}b).
The relative fraction of these fairly massive \textit{young} quiescent galaxies is therefore negligible when the stellar mass completeness limit $M_{lim}$ lies below the upturn seen around $M_* \sim 10^{9.7}  M_\odot$ (namely, when $M_{lim} \ll 10^{9.7}  M_\odot$), which is the case in the VIPERS-MLS at $z < 0.5$ with $M_{lim} \simeq 10^{8.8} M_\odot$.
Conversely, low-mass [$M_* < 10^{9.7}  M_\odot$] galaxies are mostly \textit{young} at $z < 0.5$ (Fig. \ref{fig_age_mass_sel}c).
In other words, on average, \textit{young} galaxies are \textit{low-mass} galaxies, and conversely.
However, at higher redshift, our stellar mass completeness limit reaches $M_{lim} \simeq 10^{9.4} M_\odot$ at $z < 0.65$ due to the Malmquist bias. The fraction of fairly massive [$M_* > 10^{9.7}  M_\odot$] galaxies among \textit{young} galaxies and, conversely, the fraction of \textit{old} galaxies among low-mass [$M_* < 10^{9.7}  M_\odot$] galaxies then become non negligible. We therefore have to take this into account if we want to focus on low-mass galaxies that are prone to fast quenching at $z > 0.5$.

Aiming to push our analysis to $z \sim 0.65$ (see Sect. \ref{subsect_env_z}) while ensuring a simultaneous focus on (1) \textit{young} low-mass galaxies, whose fast quenching is expected to be responsible for the low-mass upturn observed in the SMF of quiescent galaxies, and (2) \textit{old} high-mass galaxies, whose slow quenching provides the bulk of the quiescent population around $\mathcal{M}^\star$, we selected galaxies by combining $\rKs$ and stellar mass.
As illustrated in Fig. \ref{fig_age_mass_sel}d, our refined sample of low-mass ($\ylm$) galaxies was therefore selected with
\begin{equation}
[ \ \rKs < 0.76 \ ]  \  \cap \ [ M_{lim}\leq \ M_* \leq 10^{9.7}  M_\odot \ ]
\label{eq_ylm}
\end{equation}
and high-mass ($\om$) galaxies with
\begin{equation}
[ \ \rKs > 0.76  \ ] \ \cap \ [ \ 10^{9.7}  M_\odot < M_* \leq 10^{11.5} M_\odot \ ] \ ,
\label{eq_om}
\end{equation}
which ensured the low-mass galaxies we considered to be mostly \textit{young} (i.e., prone to fast quenching), even at $0.5 < z < 0.65$ (where $M_{lim} \simeq 10^{9.4} M_\odot$).

We also selected a sample of ultra-massive ($\um$) galaxies with 
\begin{equation}
10^{11.5} M_\odot < M_* \leq 10^{12} M_\odot \ ,
\label{eq_um}
\end{equation}
these galaxies being also essentially \textit{old} to $z = 0.65$ (Fig. \ref{fig_age_mass_sel}c).
This later stellar-mass bin with $M_* > 10^{11.5} M_\odot$ was motivated by the fact that these ultra-massive galaxies seem to be characterised by a peculiar evolution of their number density since $z \sim 1$ with respect to less massive galaxies \citep[see][Fig. 15]{Moutard2016b}. Being all \textit{old} galaxies and (almost) all quiescent since this epoch, their evolution is expected to be mostly driven by dry mergers in rich environments. 
Moreover, these ultra-massive galaxies embody a very advanced stage of galaxy stellar-mass assembly and set therefore a benchmark that is relevant to compare with when considering less massive galaxies.

Combining with the selection of quiescent and star-forming galaxies allowed by the NUVrK diagram (Eqs. \ref{eq_Qsel} and \ref{eq_SFsel}), this led to defining five classes of galaxies: \\
\textit{i)} quiescent (\textit{young}) low-mass galaxies, quoted Q$_\ylm$, responsible for the upturn observed in the quiescent SMF and expected to have experienced a fast quenching and \\
\textit{ii)} star-forming (\textit{young}) low-mass galaxies, quoted SF$_\ylm$, constituting the reservoir of galaxies that might experience such fast quenching; \\
\textit{iii)} quiescent (\textit{old}) high-mass galaxies, quoted Q$_\om$, responsible for the build-up of the quiescent SMF around $\mathcal{M}^\star _{\textsc{sf}} \sim 10^{10.64} M_\odot$ and expected to follow a slow quenching channel and \\
\textit{iv)} star-forming (\textit{old}) high-mass galaxies, quoted SF$_\om$, that are prone to follow this slow quenching channel; \\
\textit{v)} and finally ultra-massive galaxies, quoted Q$_\um$, already \textit{old} and mostly quiescent (in the redshift range we considered).


\section{Results}
\label{sect_results}

\subsection{Environments vs. quenching channels}
\label{sect_quench_env_relation}

Aiming to explore the impact of environment on the quenching of star formation, we compared the probability distribution functions (PDFs) of the local density $\varrho$ measured for the different categories of galaxies defined in Sect. \ref{sec_sample_sel}.

\begin{figure*}
\includegraphics[width=0.49\hsize]{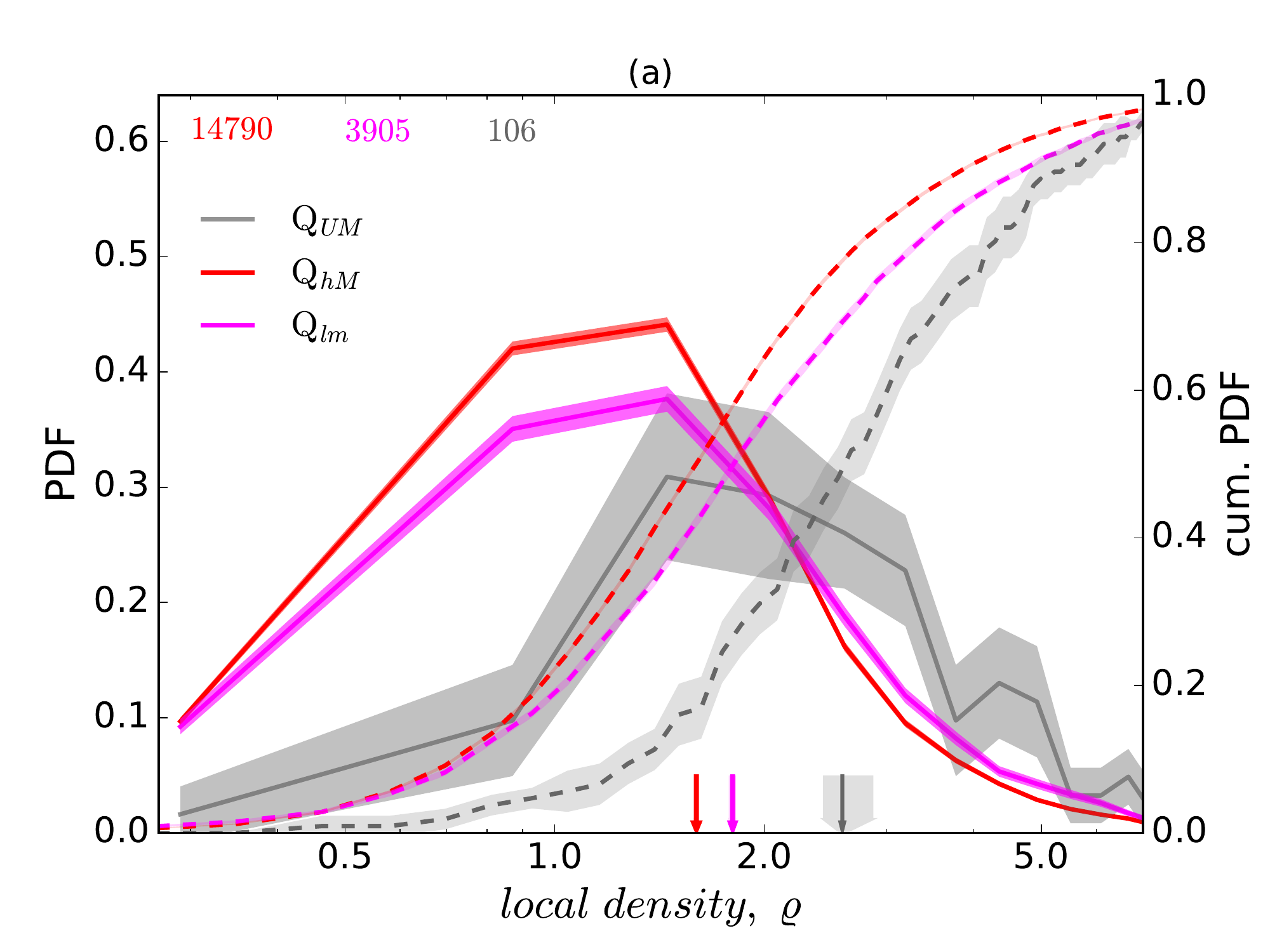}
\includegraphics[width=0.49\hsize]{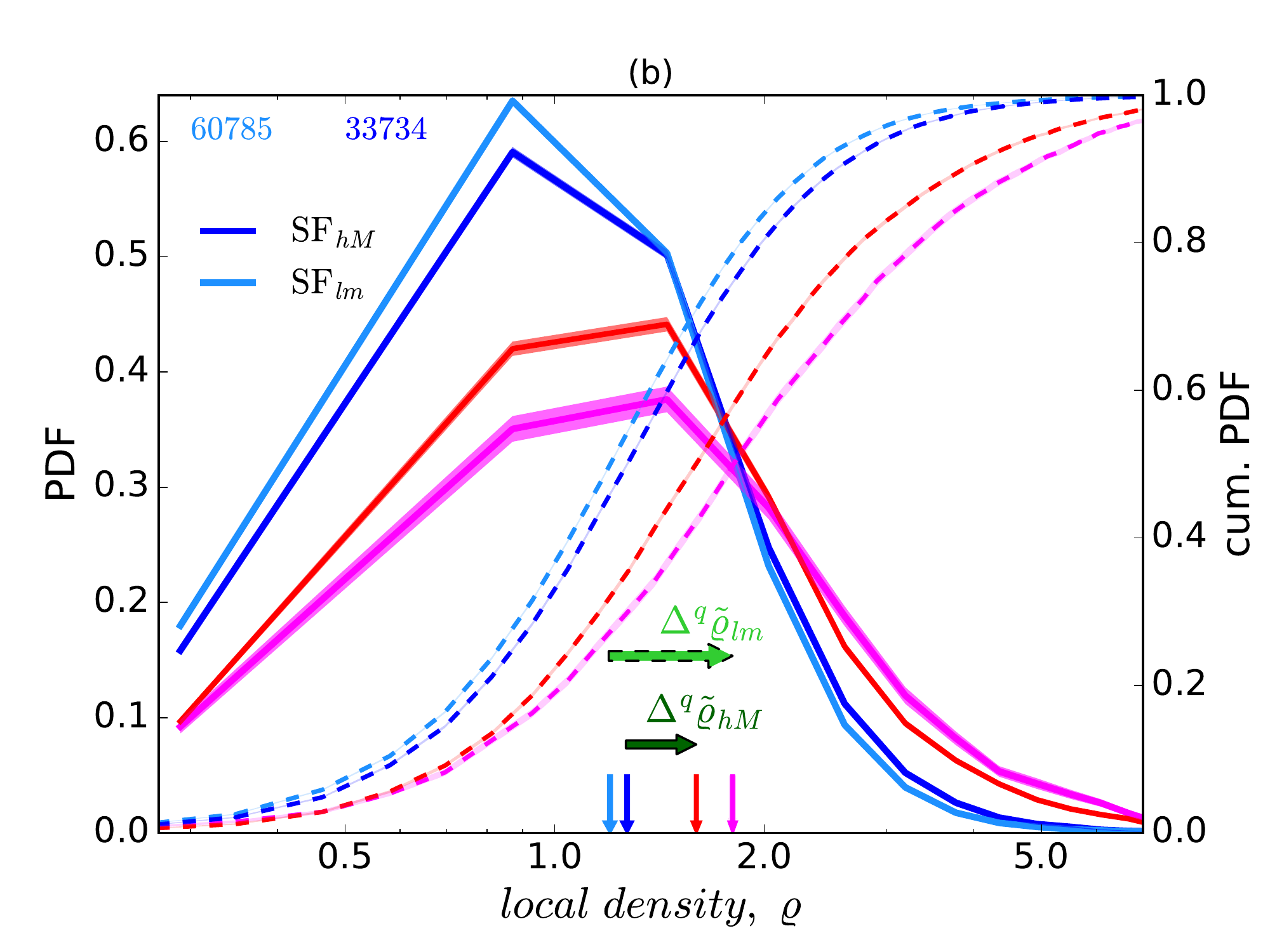}
\caption{Probability distribution function (PDF) of the local density $\varrho$, as measured in 0.5 Mpc radius apertures at redshift $0.2 < z < 0.5$ for \textbf{(a)} quiescent low-mass (magenta), high-mass (red) and ultra-massive (grey) galaxies and \textbf{(b)} comparison with star-forming counterparts for low-mass (cyan), high-mass (blue) galaxies. Dashed lines show the corresponding cumulative PDFs, while vertical arrows show the corresponding typical local densities, $\tilde{\varrho}$, defined as PDF medians. Only galaxies with $M_* \geq M_{lim}(z<0.5) = 10^{8.8} M_\odot$ are considered.
Shaded envelops represent the corresponding $\pm1\sigma$ uncertainties derived from 5000 bootstrap resamples, while the galaxy number of each subsample is written in the upper left corner. 
Horizontal green arrows (in panel b) show the local density deviation associated with the different quenching channels: 
$\Delta^q \tilde{\varrho}_\ylm$ (dashed light green arrows) and $\Delta^q \tilde{\varrho}_\om$ (dark green arrows) for low-mass galaxies prone to fast quenching and high-mass galaxies prone slow quenching, respectively (cf. Sect. \ref{sec_sample_sel}).
\label{fig_dens_QSF}  }
\end{figure*}

In Fig. \ref{fig_dens_QSF}a,
we focus on the quiescent population, divided into low-mass, high-mass and ultra-massive galaxies
at $0.2 < z < 0.5$.
In order to characterise each PDF with one single value, that can be seen as the typical local density associated with the corresponding underlying population, we computed the median for each PDF($\varrho$), denoted $\tilde{\varrho}$ (shown with vertical arrows).
The PDF and PDF-median errorbars were estimated through bootstrap resampling (made of 5000 resamples), which accounts for Poissonian uncertainties.
The first result springing up from the analysis is the confirmation that
ultra-massive galaxies clearly reside in the densest environments, with a median local density found to be $\tilde{\varrho}^\textsc{q}_{\um}=2.59^{+0.28}_{-0.16}$ (grey arrows in Fig. \ref{fig_dens_QSF}a). 
Unsurprisingly, less massive quiescent galaxies are characterised by lower local densities.

Interestingly, however, low-mass quiescent galaxies clearly appear to be located in denser environments than high-mass ones, especially at $0.2 < z < 0.5$ where $\tilde{\varrho}^\textsc{q}_{\ylm}=1.80^{+0.02}_{-0.02}$ (magenta) and $\tilde{\varrho}^\textsc{q}_{\om}=1.60^{+0.01}_{-0.01}$ (red). 
We verified that the distributions were not drown from an identical underlying  population (with respect to the projected local density) with the popular and widely used K-S test and the more robust Anderson-Darling test \citep{AndersonDarling1952, AndersonDarling1954}.\footnote{The Anderson-Darling test is notably more sensitive to the tails of the distributions \citep[for more details on statistical tests for astronomy and comparisons see, e.g.,][]{BabuFeigelson2006, Hou2009}.} Both tests confirmed that the two distributions are different at more than 99.99 per cent confidence.

Next, aiming to see if and how the quiescence of galaxies correlates with environment, we compared the PDFs of the local densities measured for quiescent (Q) galaxies with those of star-forming (SF) ones in Fig. \ref{fig_dens_QSF}b.
Similarly to Fig. \ref{fig_dens_QSF}a, local density PDFs are shown for low-mass galaxies (light blue) and high-mass ones (dark blue).\footnote{Ultra-massive galaxies are almost all quiescent, only 4/110 galaxies could have been classified as star-forming at $0.2 < z < 0.5$ (and 7/55 at $0.5 < z < 0.65$), which would prevent us from computing any robust local density PDF.} 
One can first notice a weak --though detectable-- trend that tends to make us observe high-mass star-forming galaxies in slightly denser environments than less massive ones.
At the same time, star-forming galaxies exhibit local densities that peak near the average density of the Universe, clearly less dense than that of quiescent galaxies regardless of the stellar mass. 

As a matter of fact, the difference of median density exhibited by low-mass galaxies according to the star-formation activity is unambiguously the strongest.
We have indeed $\tilde{\varrho}_{\ylm}^\textsc{sf}=1.20^{+0.01}_{-0.01}$ for star-forming galaxies and $\tilde{\varrho}_{\ylm}^\textsc{q}=1.80^{+0.02}_{-0.02}$ for quiescent ones at $0.2 < z < 0.5$. 
In fact, we may define the deviation of local density associated with quiescence, $\Delta^q \tilde{\varrho} = \tilde{\varrho}^\textsc{q} - \tilde{\varrho}^\textsc{sf}$, which represents a deviation of $\Delta^q \tilde{\varrho}_{\ylm} \simeq +0.60$ in the case of low-mass galaxies (dashed light green arrow in Fig. \ref{fig_dens_QSF}b).
As for high-mass galaxies, we also report a non-negligible but weaker local density deviation between star-forming and quiescent galaxies, with $\tilde{\varrho}_{\om}^\textsc{sf}=1.27^{+0.01}_{-0.01}$ and $\tilde{\varrho}_{\om}^\textsc{q}=1.60^{+0.01}_{-0.01}$, i.e. $\Delta^q \tilde{\varrho}_{\om} \simeq +0.33$ (dark green arrow in Fig. \ref{fig_dens_QSF}b).

This confirms that quiescent galaxies are, on average, located in denser environments than star-forming ones, both for low-mass and high-mass galaxies.
But the key finding of our analysis is that low-mass galaxies, also identified as prone to being quenched through a fast quenching channel (cf. Sect. \ref{sec_sample_sel}), require a much stronger increase of their typical local density to be observed as quiescent than high-mass galaxies, prone to follow a slow quenching channel, as we discuss later in this paper (Sect. \ref{subsec_env_quench_lm_gal}).


\subsection{Local density evolution}
\label{subsect_env_z}

Aiming to observe the evolution of typical densities across redshift, we considered the additional redshift bin $0.5 < z < 0.65$. 
The upper redshift limit was set so that it allowed us to probe the excess of low-mass quiescent galaxies at higher redshift while being complete in mass, in addition to ensuring two redshift bins of similar comoving volumes.\footnote{Respectively, 
$\sim 4.97 \times 10^6$ $h^{-3}$ Mpc$^3$ and $\sim 4.84 \times 10^6$ $h^{-3}$ Mpc$^3$ at $0.2 < z < 0.5$ and $0.5 < z < 0.65$.}
To enable comparison of local densities between our redshift bins, we repeated the same analysis than what is presented in Sect. \ref{sect_quench_env_relation}, but we only considered galaxies more massive than the stellar mass completeness limit of the highest redshift bin, namely $M_{lim} = 10^{9.4} M_\odot$, both at $0.2 < z < 0.5$ and $0.5 < z < 0.65$.

\begin{figure*}
\includegraphics[width=\hsize]{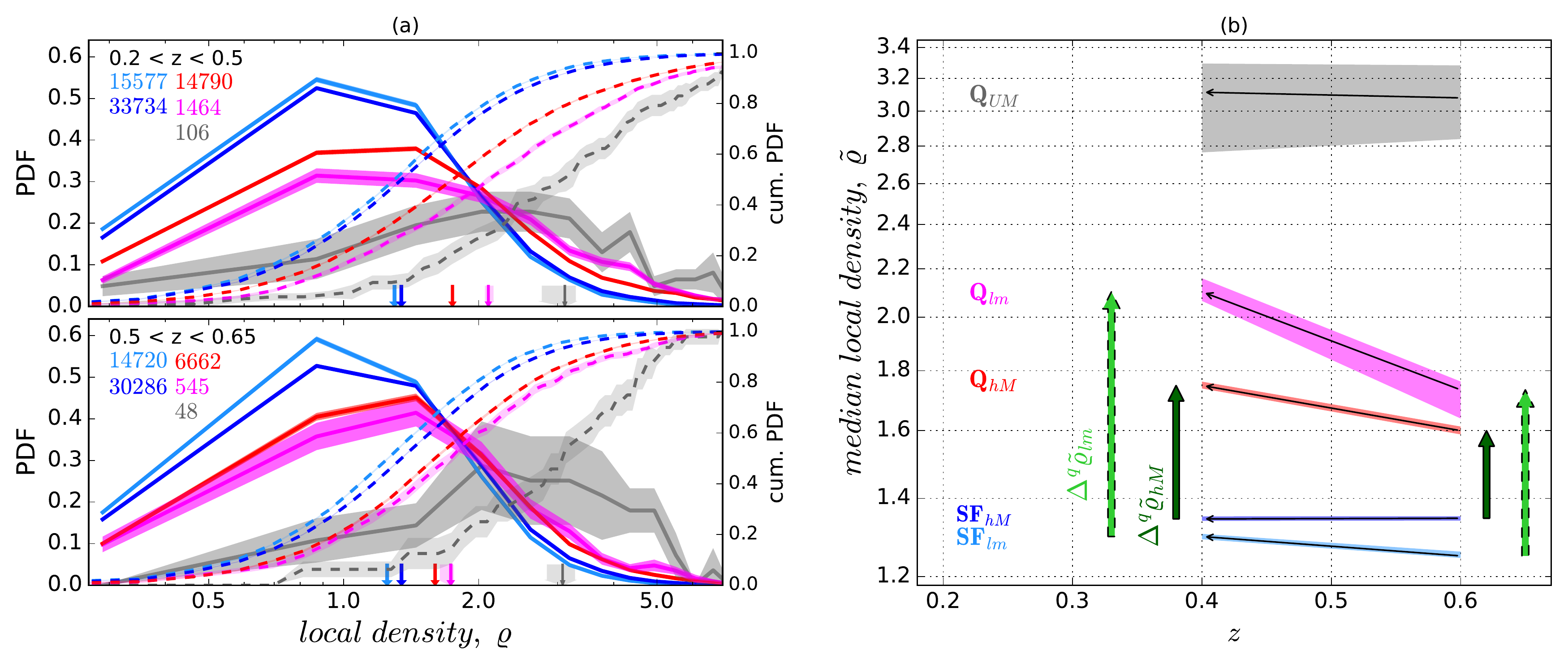}
\caption{Evolution of local densities with redshift, considering galaxies with $M_* \geq M_{lim}(z<0.65) = 10^{9.4} M_\odot$. \textbf{(a)} PDF($\varrho$) (solid lines) and cumulative PDF($\varrho$) (dashed lines) at redshift $0.2 < z < 0.5$ (top) and $0.5 < z < 0.65$ (bottom) for the different classes of galaxies defined in Sect. \ref{sec_sample_sel}: low-mass quiescent ($\mathrm{Q}_{\ylm}$; magenta) and star-forming ($\mathrm{SF}_{\ylm}$; light blue) galaxies, high-mass quiescent ($\mathrm{Q}_{\om}$; red) and  star-forming ($\mathrm{SF}_{\om}$; dark blue) ones and ultra-massive galaxies ($\mathrm{Q}_{\um}$; grey). The galaxy number of each subsample is written in the upper left corner.
\textbf{(b)} Corresponding evolution of the median local density $\tilde{\varrho}$ (cf. Sect. \ref{sect_results}) with redshift between $z \sim 0.6$ and $z \sim 0.4$ (median redshifts of $0.5 < z < 0.65$ and $0.2 < z < 0.5$, respectively).
Similarly to Fig. \ref{fig_dens_QSF}, shaded envelopes represent the corresponding $\pm1\sigma$, as derived from bootstrap resampling.
Vertical green arrows show the local density deviation associated with the different quenching channels, at $0.2 < z < 0.5$ and $0.5 < z < 0.65$; 
$\Delta^q \tilde{\varrho}_\ylm$ (dashed light green arrows) and $\Delta^q \tilde{\varrho}_\om$ (dark green arrows) for low-mass galaxies prone to fast quenching and high-mass galaxies prone slow quenching, respectively (cf. Sect. \ref{sec_sample_sel}).
 \label{fig_dens_QSF_z}  }
\end{figure*}


In Fig. \ref{fig_dens_QSF_z}a we show PDF($\varrho$) for the different classes of galaxies we selected (cf. Sect. \ref{sec_sample_sel}), as traced by galaxies with $M_* > 10^{9.4} M_\odot$ both at $0.2 < z < 0.5$ (top) and $0.5 < z < 0.65$ (bottom), where vertical arrows reflect the corresponding values of $\tilde{\varrho}$, similarly to Fig. \ref{fig_dens_QSF}. 
One may notice how the introduction of a higher stellar-mass completeness limit ($M_{lim} = 10^{9.4} M_\odot$ instead of $M_{lim} = 10^{8.8} M_\odot$) reduces the number of low-mass galaxies at $0.2 < z < 0.5$ (compared to Fig. \ref{fig_dens_QSF}), galaxies with $M_{lim} < 10^{9.4} M_\odot$ being discarded, and consequently how this affects the measurements of the local density $\varrho$ (as traced by the median local density $\tilde{\varrho}$). The total number of galaxies is indeed reduced when considering $M_* > M_{lim} = 10^{9.4} M_\odot$ instead of $M_* > M_{lim} = 10^{8.8} M_\odot$ and we recall that $\varrho$ is normalized by the mean density of the Universe at the same redshift (cf. Sect. \ref{sect_dens_measur}), which proportional to the total number of considered galaxies (Eq. \ref{eq_loc_dens}). It is therefore expected to measure lower values of $\tilde{\varrho}$ when considering a higher stellar-mass limit.
The trends we measure are, however, consistent at $z < 0.5$ with $M_{lim}(z<0.5) = 10^{8.8} M_\odot$ or $M_{lim}(z<0.65) = 10^{9.4} M_\odot$, which confirms our conclusions.

One may thus notice that, as at $0.2 < z < 0.5$, ultra-massive galaxies are characterized by the highest local density we measured and that, at lower stellar mass, quiescent galaxies are generally characterised by much higher local local densities than star-forming ones at $0.5 < z < 0.65$. 
In particular, considering quiescent galaxies, Q$_\ylm$ galaxies may already be characterized by higher local densities than Q$_\om$ ones at $0.5 < z < 0.65$, although considering PDF uncertainties, local densities of low-mass and high-mass quiescent galaxies may be considered to be similar.\footnote{At $0.5 < z < 0.65$, the confidence given by an Anderson-Darling test of having $\tilde{\varrho}_{\ylm}$ > $\tilde{\varrho}_{\om}$ is 99.99 per cent for median PDFs, 93 per cent when considering the $\pm1\sigma$ PDF uncertainties estimated from 5000 bootstrap resamples (i.e., confidence level is at least of 93 per cent for 84 per cent of resamples; cf. envelops reported in Fig. \ref{fig_dens_QSF_z}) and this confidence would drop to 57 per cent in a very conservative approach considering $\pm2\sigma$ PDF uncertainties (i.e., confidence level is $> 57$ per cent for 95.6 per cent of resamples).}
In any event, the deviation observed between the local density of quiescent and star-forming galaxies for low-mass galaxies ($\Delta^q \tilde{\varrho}_{\ylm} \simeq +0.48$) is already larger than for high-mass galaxies ($\Delta^q \tilde{\varrho}_{\om} \simeq +0.35$) at $0.5 < z < 0.65$.
The fact that the quiescence of low-mass galaxies, associated with fast quenching (cf. Sect. \ref{sec_sample_sel}), 
requires a stronger increase of the local density to be observed than the quiescence of high-mass galaxies is therefore confirmed at $0.5 < z < 0.65$ as well.


Focussing on the redshift evolution of the typical local density, Fig. \ref{fig_dens_QSF_z}b shows the evolution of $\tilde{\varrho}$ between $z \sim 0.6$ (median redshift of galaxies in our highest redshift bin, $0.5 < z < 0.65$) and $z \sim 0.4$ (median redshift of galaxies at $0.2 < z < 0.5$) for the different classes of galaxies we considered.
It is thus interesting to notice how constant the typical local density of high-mass star-forming galaxies appears to be constant ($\Delta^z \tilde{\varrho}_{\om}^\textsc{sf} \simeq 0$), whilst the typical local density of their quiescent counterparts is characterised by a clear increase of $\Delta^z \tilde{\varrho}_{\om}^\textsc{q} \simeq +0.15$ between $z \sim 0.6$ and $z \sim 0.4$. This results in the increase of the local density deviation between star-forming and quiescent high-mass galaxies from $\Delta^q \tilde{\varrho}_{\om}\simeq +0.25$ at $z \sim 0.6$ to $\Delta^q \tilde{\varrho}_{\om}\simeq +0.40$ at $z \sim 0.4$, therefore essentially due to the fact that the local density of high-mass quiescent galaxies has increased with cosmic time, on average.

At the same time, Fig. \ref{fig_dens_QSF_z}b reveals an even stronger increase of the typical local density $\tilde{\varrho}_{\ylm}$ for low-mass quiescent galaxies, with a variation of $\Delta^z \tilde{\varrho}_{\ylm}^\textsc{q} \simeq +0.52$, which needs to be weighted by the fact that their star-forming counterparts also experienced a small increase of their typical local density with $\Delta^z \tilde{\varrho}_{\ylm}^\textsc{sf} \simeq +0.05$. 
Still, this results in what appears to be a strong increase of local density deviation observed between star-forming and quiescent for low-mass galaxies from $z \sim 0.6$ to $z \sim 0.4$, namely, from  $\Delta^q \tilde{\varrho}_{\ylm}\simeq +0.48$ to $\Delta^q \tilde{\varrho}_{\ylm}\simeq +0.80$, compared to high-mass galaxies. 
While this traces indeed the fact that, on average, the local density of low-mass quiescent galaxies has increased faster with cosmic time than what we observe for high mass galaxies, we will see how the remarkable increases of both $\tilde{\varrho}_{\ylm}^\textsc{q}$ (i.e., $\Delta^z \tilde{\varrho}_{\ylm}^\textsc{q}$ by definition) and $\Delta^q \tilde{\varrho}_{\ylm}$ with cosmic time may be explained by the simultaneous modest increase of $\tilde{\varrho}_{\ylm}^\textsc{sf}$ (see Sect. \ref{subsec_rising_env_quench}).

One may finally notice that $\tilde{\varrho}^\textsc{q}_{\um}$ might exhibit a very small increase with cosmic time, with $\Delta^z \tilde{\varrho}^\textsc{q}_{\um} \simeq +0.03$ between $z \sim 0.6$ and $z \sim 0.4$ (from $\tilde{\varrho}^\textsc{q}_{\um} = 3.08^{+0.20}_{-0.24}$ to $\tilde{\varrho}^\textsc{q}_{\um} = 3.11^{+0.18}_{-0.35}$). However, one can see how the uncertainties affecting $\tilde{\varrho}^\textsc{q}_{\um}$ allow for a variation  $-0.52 \lesssim \Delta^z \tilde{\varrho}^\textsc{q}_{\um} \lesssim +0.45$, which prevents us from drawing any conclusion about the local density evolution experienced by ultra-massive galaxies.

\section{Discussion}
\label{sect_discuss}

We have seen in Sect. \ref{sect_results} how different may be the local density of galaxies depending on whether they are quiescent or not and, above all, depending on the quenching channel they are prone to follow (fast for low-mass galaxies or slow for high-mass galaxies), and then how this may evolve with cosmic time at $0.2 < z < 0.65$.
In this section, we discuss our results and notably the connection between environment and star-formation quenching that may be highlighted, in particular, the impact of environment on the (fast) quenching of (\textit{young}) low-mass galaxies and its evolution with cosmic time.

\subsection{Ultra-massive galaxies reside in very dense environments}

As is obvious in Fig. \ref{fig_dens_QSF_z}b, ultra-massive galaxies are by far located in the densest environments that were measured in our analysis.
These ultra-massive galaxies are almost all quiescent and characterised by old stellar populations from $z \sim 0.6$.
This makes them good candidates for subsequent growth via (dry) mergers, as already proposed \citep[see, e.g.,][]{DeLucia2006, DeLucia2007, Cattaneo2011, Moutard2016b, Lee2017, Groenewald2017}.

At the same time, though non-negligible compared to smaller surveys at the similar redshift, the limited number of ultra-massive ($M_* > 10^{11.5} M_\odot$) galaxies in our analysis (106 at $0.2 < z < 0.5$, 48 at $0.5 < z < 0.65$) prevented us from constraining the evolution of their local density at $0.2 < z < 0.65$. 
Constraining such evolution would be of high interest to explore the growth of structures on different scales. Indeed, for instance, a decreasing local density around an ultra-massive galaxy may support a picture where the galaxy merger rate within the host structure is higher than the rate at which new galaxies fall onto the structure, and vice versa.

In any case, the high local densities measured around ultra-massive galaxies support a picture where these galaxies are experiencing a very advanced stage of both galaxy stellar-mass assembly and galaxy clustering.

\subsection{The role of environment in the quenching of low-mass galaxies}
\label{subsec_env_quench_lm_gal}

As described in Sect. \ref{sect_results}, when focussing on high-mass galaxies, one can see that quiescent galaxies are characterised by higher typical local densities than star-forming ones, as traced by $\Delta^q \tilde{\varrho}_\om$, both at $0.2 < z < 0.5$ and $0.5 < z < 0.65$ (Fig. \ref{fig_dens_QSF}b).
This is expected because among high-mass star-forming galaxies, the most massive quench first \citep[see, e.g.,][]{Moutard2016b}. At the same time, more massive galaxies are expected to be more clustered on large-scales \citep[typically, what happens around filaments;][]{Malavasi2017}. More massive galaxies are indeed hosted by more massive DM halos, on average, while halo clustering increases with halo mass (given the hierarchical growth of DM structures with cosmic time). 
In this respect, our study is therefore consistent with many previous studies that have emphasized the fact that quiescent galaxies are preferentially located in denser environments, and in particular concerning massive galaxies with $M_* \gtrsim 10^{10} M_\odot$ \citep[e.g.,][]{Kauffmann2004, Baldry2006, Lani2013, Malavasi2017, Etherington2017, Cucciati2017}. 

The interest of the present analysis is, however, its ability to disentangle the impact of environment on different categories of galaxies that are prone to follow different quenching channels: slowly quenched (\textit{old}) high-mass galaxies feeding the quiescent population around $\mathcal{M}^\star _{\textsc{sf}} \simeq 10^{10.64} M_\odot$, and (\textit{young}) low-mass galaxies subject to a fast quenching (cf. Fig. \ref{fig_age_mass_sel}) responsible for the excess of quiescent galaxies at $M_* < 10^{9.7}  M_\odot$.
Thus, the first remarkable result of our analysis is the fact that these low-mass quiescent galaxies were already located in denser environments than high-mass quiescent galaxies at $0.2 < z < 0.5$ and probably as of $0.5 < z < 0.65$, as observed in the local Universe \citep[e.g.,][]{Hogg2003, Haines2007}.

The role of environment in the quenching of low-mass galaxies is confirmed by the deviation of the typical local density observed between the star-forming and quiescent populations, $\Delta^q \tilde{\varrho}$: besides the fact that quiescent low-mass galaxies appear to be located in much denser environment than their star-forming counterparts, the local density deviation between star-forming and quiescent galaxies is more than twice stronger for low-mass galaxies ($\Delta^q \tilde{\varrho}_{\ylm} \simeq +0.80$) than for high-mass ones ($\Delta^q \tilde{\varrho}_{\om} \simeq +0.41$). 
In other words, the quiescence of low-mass galaxies requires a much stronger increase of the local density than the quiescence of high-mass galaxies.

This is therefore consistent with a picture where the upturn observed at low-mass in the SMF of quiescent galaxies is due to the (fast) quenching of (\textit{young}) low-mass galaxies, due to mechanisms that involve rich environments, as observed in the local Universe \citep{Hogg2003, Haines2007, Peng2012}.
Our analysis shows that such a picture is also valid at $0.2 < z < 0.65$, confirming and complementing the study of \citet{Guo2017} who recently correlated the quenching of low-mass galaxies with rich environments at $0.5 < z < 1$ in the CANDELS fields.
While confirming that environment already played a significant role at earlier times, when large-scale structures were less developed, this raises the question of the importance of \textit{environmental quenching} across cosmic time.

\subsection{A rising importance of environmental quenching with cosmic time?}
\label{subsec_rising_env_quench}

When focussing on low-mass galaxies, we noticed in Sect. \ref{subsect_env_z} that the typical local density of SF galaxies slightly increased from $z \sim 0.6$ to $z \sim 0.4$\footnote{We recall that, to ensure completeness over the entire redshift range $0.2 < z < 0.65$, low-mass galaxies have stellar masses $10^{9.4} M_\odot \leq M_* < 10^{9.7}  M_\odot$.} ($\Delta^z \tilde{\varrho}_{\ylm}^\textsc{sf}$ in Fig. \ref{fig_dens_QSF_z}b), which reflects the fact that an increasing fraction of SF$_\ylm$ galaxies has been characterised by rich environments with decreasing redshift.
At the same time, we observed a stronger increase of the typical density for Q$_\ylm$ galaxies, even already found to be located in much richer environments than their SF$_\ylm$ counterparts from $z \sim 0.6$ ($\Delta^z \tilde{\varrho}_{\ylm}^\textsc{q}$ in Fig. \ref{fig_dens_QSF_z}b).
While the increase of the typical local density is due to the growth of large-scale structures that host a growing number of galaxies, the fairly modest increase of $\Delta^z \tilde{\varrho}_{\ylm}^\textsc{sf}$ observed for SF$_\ylm$ galaxies is expected if the vast majority of these galaxies are field galaxies, while Q$_\ylm$ galaxies are preferentially located in rich environments and fully experience the growing number of galaxies within large-scale structures.

Actually, while the number of SF$_\ylm$ galaxies located in fairly rich environments is small compared to the total number of SF$_\ylm$ galaxies, it represents a significant number compared to the number of corresponding Q$_\ylm$ galaxies. 
The size of the entire SF$_\ylm$ galaxy population is indeed 10--30 times larger than that of Q$_\ylm$ galaxies in our sample (cf. Fig \ref{fig_dens_QSF_z}). 
For example, the fraction of SF$_\ylm$ galaxies located in very dense regions where we measure $\varrho > 4$ (i.e., 4 times the mean local density of the Universe at the same redshift) is only 1.3 per cent at $z \sim 0.6$ and 2.7 per cent at $z \sim 0.4$, which represents an increase of the number of these galaxies from 195 to 435, whilst the corresponding fraction of Q$_\ylm$ galaxies increased from 9.5 per cent to 19.6 per cent, but involving fewer galaxies, with 52 and 287 Q$_\ylm$ galaxies at $z \sim 0.6$ and $z \sim 0.4$, respectively.
This highlights how the increasing number of SF$_\ylm$ galaxies that are characterised by very rich environments is able to feed the strong increase of the number of Q$_\ylm$ observed in corresponding environments (the contribution of Q$_\ylm$ to the low-mass population increasing from 21 per cent to 40 per cent between $z \sim 0.6$ and $z \sim 0.4$).
It is, moreover, interesting to note here that SF$_\ylm$ galaxies remain more numerous than their Q$_\ylm$ counterparts (79 per cent to 60 per cent at  $z \sim 0.6$ and $z \sim 0.4$, respectively) in these very dense regions, as discussed in the next section.
In other words, this tends to confirm a picture where the reservoir of low-mass galaxies susceptible to environmental quenching is growing with cosmic time, following the growth of large-scale structures that host a growing number of galaxies.

On the other hand, the fact that the comoving number density of low-mass quenched galaxies has increased with cosmic time does not mean that the corresponding quenching has become more important: the number of low-mass galaxies having quenched via environmental quenching has to be compared with that of high-mass galaxies quenched via mass quenching across cosmic time.
In order to quantify the contribution of the environmental quenching channel followed by low-mass galaxies, one may define the \textit{low-to-high-mass ratio} of the quiescent population at given redshift, $\mathcal{R}^\textsc{q}_{\ylm/\om}$,  derived as the comoving number density of low-mass environmentally-quenched galaxies $\mathcal{N}_{\ylm}^\textsc{q}$ relative to that of high-mass mass-quenched galaxies $\mathcal{N}_{\om}^\textsc{q}$, as
\begin{equation}
\mathcal{R}^\textsc{q}_{\ylm/\om} =  \frac{\mathcal{N}_{\ylm}^\textsc{q}}{\mathcal{N}_{\om}^\textsc{q}} \ .
\end{equation}
One may thus observe a modest but detectable increase of this ratio from $\mathcal{R}^\textsc{q}_{\ylm/\om} = 0.082\pm0.005$ to $0.099\pm0.003$ between $z \sim 0.6$ and $z \sim 0.4$. 
Yet, this seeming evolution of $\mathcal{R}^\textsc{q}_{\ylm/\om}$ might be artificial, due to the fact that faint quiescent galaxies are expected to be the firsts to suffer from incompleteness with increasing redshift.
As a matter of fact, the completeness limit we adopted (namely, $M_* \geq M_{lim} = 10^{9.4} M_\odot$) ensures our quiescent sample to be more than 95 per cent complete at $z < 0.65$, but in the particular case of low-mass quiescent galaxies, the completeness can drop to $\sim80$ per cent around $z \sim 0.6$ (against $\geq95$ per cent at $z < 0.5$). If we assume, in a conservative approach, that all low-mass galaxies suffer from such incompleteness at $z > 0.5$, the low-to-high-mass ratio of the quiescent population would rather approach $\mathcal{R}^\textsc{q}_{\ylm/\om} = 0.097\pm0.005$ at $z \sim 0.6$ (against $0.099\pm0.003$ at $z \sim 0.4$),
which would therefore be consistent with no evolution of $\mathcal{R}^\textsc{q}_{\ylm/\om}$ with cosmic time at $0.2 < z < 0.65$. 
In other words, the differential incompleteness of low-mass quiescent galaxies at $z < 0.5$ and $z < 0.65$ might be sufficient to explain the increase of $\mathcal{R}^\textsc{q}_{\ylm/\om}$ that we detected between $z \sim 0.6$ and $z \sim 0.4$.

Nevertheless, the rapid build-up of the low-mass quiescent population observed over the same redshift range from deeper surveys \citep[$\sim$0.5 dex around $M_* \sim 10^{9} M_\odot$, against $\sim$0.1 dex around $M_* \sim 10^{10.5} M_\odot$, e.g., in COSMOS;][]{Davidzon2017} suggests a rising share of low-mass galaxies in the quiescent population, which might plead for a rising importance of the environmental-quenching channel (followed by low-mass galaxies) compared to the mass-quenching channel (followed by high-mass galaxies). 
This picture might be supported by the fact that the highest density regions reveal a rising fraction of low-mass quiescent galaxies with cosmic time from $z \sim 2$ \citep[e.g.,][]{Papovich2018}, but the corresponding number of environmentally-quenched galaxies should be compared to the simultaneous number of mass-quenched galaxies. 
Upcoming large surveys combining deeper optical and near-infrared observations will allow us to verify whether the importance of the environmental quenching channel followed by low-mass galaxies has risen with cosmic time at late epochs.

In any case, our results confirmed that a rising number of low-mass galaxies have been prone to experiencing environmental quenching with cosmic time.
The mechanism(s) that may be involved in such environment-driven quenching of low-mass galaxies remain(s), however, a matter of debate, which might be interesting to address in the light of all the elements we gathered so far.

\subsection{Composite picture of the environmental quenching channel followed by low-mass galaxies}

As discussed extensively in the present paper, the quenching of low-mass galaxies is associated with a strong increase of their local density, which allow us to link the quenching of these galaxies with environmental effects.


At the same time, low-mass quiescent galaxies have been shown to be essentially \textit{young} quiescent galaxies (i.e., characterised by \textit{young} stellar populations; cf. Sect. \ref{sec_sample_sel}), which has been shown to require a fast quenching \citep[see, e.g.,][]{Schawinski2014, Moutard2016b, Pacifici2016a, Pacifici2016b}. Low-mass quiescent galaxies are therefore recently quenched galaxies.
This is consistent with the fact that they exhibit rest-frame colours that are similar to those of post-starburst galaxies \citep{Kriek2010, Whitaker2012},\footnote{Our \textit{young} quiescent population, selected in the NUVrK diagram with rest-frame colours r--$\Ks$ < 0.76, overlaps at more than 87 per cent with a sample of the \textit{young} quiescent galaxies selected in the UVJ diagram with rest-frame colours U--V < 0.9 by \citet{Whitaker2012} as post-starburst galaxies.}
the incidence of which is found to be enhanced in very rich environments \citep[e.g.,][]{Paccagnella2017, Socolovsky2018}.

It has also been claimed that \textit{dwarf} satellite galaxies (corresponding to our low-mass galaxies\footnote{We verified that \textit{dwarf} galaxies of \citet{Haines2007} and our low-mass galaxies overlap at more than 80 per cent.}) may be characterised by long quenching time-scales \citep{Haines2007}. That statement was based on the fact that a significant fraction of dwarf satellite galaxies was found to be star-forming in the local Universe, while exhibiting slightly lower star-formation rates than in their field counterparts. 
Our interpretation is, on the contrary, that those results are consistent with a fast quenching of dwarf satellite galaxies. Indeed, the H$\alpha$ equivalent-width distribution measured by \citet[][Fig. 5]{Haines2007} for dwarf galaxies has only revealed a very small number of transitioning galaxies with respect to that observed in the star-forming and quiescent sequences. And, if dwarf satellite galaxies were slowly quenched, one could expect to statistically observe a significant fraction of them in transition between the star-formation and quiescent sequences, which is not observed.

Rather than slow quenching, those results pleads for a fast quenching of dwarf satellite galaxies in the local Universe, but delayed in onset, since more than 60 per cent of them are star-forming \citep{Haines2007},
which agrees with SMF measurements for central and satellite galaxies in the local Universe where more than 50 per cent of low-mass $M_* < 10^{9.7}  M_\odot$ satellite galaxies are star-forming \citep{Yang2009, Peng2012}. 
It is interesting to note that our observations highlight a similar trend at $0.2 < z < 0.65$, where 79 per cent and 60 per cent of low-mass galaxies with high local densities ($\varrho > 4$) --i.e., prone to fast environmental quenching-- are star-forming at $z \sim 0.4$ and $z \sim 0.6$, respectively (cf. Sect. \ref{subsec_rising_env_quench}).
Indeed, \textit{delayed-then-rapid} quenching scenarii, initially proposed in the local Universe to reproduce the SFR distribution of satellite galaxies in clusters \citep[][]{Wetzel2013,Oman2016}, have recently been shown to be well suited at $0.5 \lesssim z \lesssim 1$ as well, with an increasing delay before quenching with decreasing stellar mass \citep{Fossati2017}. In such scenarii, the quenching of a satellite galaxy is expected to take a few hundred Myrs, but it occurs several Gyrs after the infall onto the group or cluster. 
However, as shown by \citet{Haines2007}, the fact that dwarf star-forming satellite galaxies exhibit slightly smaller H$\alpha$ emission (which traces almost instantaneous SFR) than their field counterparts may highlight the quenching of a part of the star-formation in low-mass galaxies upon or shortly after becoming satellites.


The picture may finally be complemented by the fact that \textit{young} quiescent galaxies have been shown to be mostly bulge-dominated\footnote{We focussed on galaxies with semi-major axis $A > 50$ pixels (i.e, $A > 8.4\arcsec$) at $z < 0.25$.} \citep[][Fig. 16]{Moutard2016a}, which implies that environmental quenching of low-mass galaxies is probably combined with a rapid morphological transformation, consistently with what has been observed in the local Universe \citep{Schawinski2014} and at higher redshift \citep[$0.5<z<1$; ][]{Kawinwanichakij2017}.
In summary, we may therefore have to consider any scenario supporting a delayed-then-rapid quenching of satellite galaxies, where star formation is suppressed in $\sim0.4$ Gyr \citep[][]{Moutard2016b} and associated with a simultaneous transformation of galaxy morphology.
For example, ram-pressure stripping processes, able to suppress star-formation of a satellite galaxy over 0.2--0.8 Gyrs when it reaches the core of a cluster 2--4 Gyrs after entering it \citep{Mahajan2011, Wetzel2013, Muzzin2014}, would require to be associated with tidal stripping harassment to alter the morphology \citep{Moore1996}. 
Alternatively, the incidence of \textit{young} low-mass quiescent galaxies in rich environments may be consistent with a major role of mergers within clusters \citep[e.g.,][]{Schawinski2014}, by nature compatible with a delayed-then-rapid quenching scenario, while being associated with almost instantaneous transformation of the morphology.

However, it has been shown that the quenching scenario may be quite different depending on the scale of the involved structures \citep[groups or clusters; e.g.,][]{Lin2014}. 
While the aim of the present study was to highlight the role of environment in the fast quenching of low-mass galaxies, the characterisation of the scale at which environmental quenching of low-mass galaxies operates will allow us to specify the physical mechanisms at play.

\section{Summary}

In an earlier paper \citep{Moutard2016b}, we identified two different quenching channels in the rest-frame NUV--r vs. r--K (i.e., NUVrK) colour diagram:
one quenching channel is followed by evolved star-forming galaxies (characterized by \textit{old} stellar populations) and is expected to be slow, while the other is required to explain the presence of \textit{young} quiescent galaxies (characterized by \textit{young} stellar populations) and is expected to be $\sim 2-9$ times faster. 

The first quenching channel is followed by high-mass galaxies, typically turning quiescent when reaching a characteristic stellar masses of $\mathcal{M}^\star_\textsc{sf} \simeq 10^{10.64} M_\odot$, which is consistent with \textit{mass quenching} \citep{Ilbert2010, Peng2010}. 
In contrast, the other quenching channel is essentially followed by low-mass [$M_* < 10^{9.7} M_\odot$] galaxies that are responsible for the upturn observed in the SMF of quiescent galaxies, which raised the question of environment role in such quenching channel: is the fast quenching of low-mass galaxies consistent with environmental quenching?
Furthermore, the rapid build-up this excess of low-mass quiescent galaxies observed from $z \sim 1$ \citep[e.g.,][]{Tomczak2014, Davidzon2017} may suggest a rising importance taken by \textit{environmental quenching} compared to \textit{mass quenching} with cosmic time 
(i.e., its rising contribution to the build-up of the quiescent population), as expected in the context of the growth of large-scale structures with cosmic time \citep[e.g.,][]{Peng2010}.

In the present paper, we analysed the relation between quenching and environment aiming, in particular, to determine the role played by environment in the quenching of low-mass galaxies.
Making use of a galaxy sample complete down to stellar masses of $M_* \geq 10^{9.4} M_\odot$ to $z \sim 0.65$ ($M_* \geq 10^{8.8} M_\odot$ to $z \sim 0.5$)  including more than 33,500 (43,000) quiescent galaxies from the VIPERS Multi-Lambda Survey \citep[VIPERS-MLS;][]{Moutard2016a}, we selected galaxies according to the quenching channel they are prone to follow in the NUVrK rest-frame colour diagram while, thanks to accurate photometric redshifts ($\sigma_{\delta z/(1+z)} < 0.04$), galaxy environment was characterized through local density measurements.
We summarise our main conclusions below.
\begin{itemize}

\item[1. ] In addition to being already mostly quiescent at $0.2 < z < 0.65$, ultra-massive [$10^{11.5} M_\odot < M_* \leq  10^{12} M_\odot$] galaxies are characterised by the highest local densities measured in our analysis.
This confirms a picture where quiescent ultra-massive galaxies may grow in mass via subsequent (dry) mergers at late epochs \citep[e.g.,][]{DeLucia2006, Cattaneo2011, Moutard2016b, Groenewald2017}. \\
High-mass [$10^{9.7}  M_\odot < M_* \leq 10^{11.5} M_\odot$] quiescent galaxies appear to be generally located in denser environments than their star-forming counterparts.
At the same time, the typical local density of high-mass star-forming galaxies appears to be constant between $z \sim 0.6$ and $z \sim 0.4$.
This is consistent with a picture where the most massive --and therefore most clustered-- among high-mass star-forming galaxies quench first \citep[e.g.,][]{Bundy2006, Ilbert2010, Davidzon2013, Moutard2016b}.

\item[2. ] Interestingly, we found that low-mass [$M_* \leq 10^{9.7}  M_\odot$] quiescent galaxies are, on average, characterized by much denser environments than high-mass quiescent galaxies at $0.2 < z < 0.5$, and probably already at  $0.5 < z < 0.65$.
Furthermore, the deviation of typical local density observed between quiescent and star-forming low-mass galaxies is always much larger than what can be observed for high-mass galaxies, both at $0.2 < z < 0.5$ and $0.5 < z < 0.65$, which implies that the quiescence of low-mass galaxies requires, on average, a much stronger increase of the local density than for high-mass galaxies. 
This highlights the lead role of environment in the fast quenching of low-mass galaxies at $0.2 < z < 0.65$, consistently with observations made in the local Universe \citep[e.g.,][]{Hogg2003, Haines2007} and recently at higher redshift \citep[namely, at $0.5 < z <1.0$;][]{Guo2017}. \\
In particular, our results confirm that environmental quenching is responsible for the low-mass upturn observed in the SMF of quiescent galaxies at $0.2 < z < 0.5$, consistently with what is observed in the local Universe \citep{Yang2009, Peng2012}.

\item[3. ] While the apparent increase of the low-mass galaxy share in the quiescent population that we observed between $z \sim 0.6$ and  $z \sim 0.4$ may confirm a rising importance taken by \textit{environmental quenching} over \textit{mass quenching} with cosmic time, this might be dominated by the differential incompleteness affecting our sample of low-mass quiescent galaxies at $z < 0.65$ and  $z < 0.5$. 
The simultaneous increase of the typical local density we measured for star-forming low-mass galaxies highlights, however, a clear growth of the reservoir of low-mass galaxies prone to environmental quenching with cosmic time at $0.2 < z < 0.65$. 
Deeper large surveys will soon allow us to confirm whether environmental quenching has become predominant in the feeding of the quiescent population at late epochs, as suggested by the rapid build-up of the SMF low-mass end for quiescent galaxies \citep[][]{Tomczak2014, Davidzon2017}.

\item[4. ] Combining our results with previous studies, 
we finally refined the composite profile of the quenching process affecting low-mass galaxies.
Namely, we have converged to a scenario consistent with the \textit{delayed-then-rapid} quenching of satellite galaxies \citep[][]{Wetzel2013}, in which low-mass galaxies would remain star-forming after entering the over-dense region to eventually experience a fast quenching in $\sim0.4$ Gyr \citep[][]{Moutard2016b} while being probably associated with a simultaneous transformation of galaxy morphology \citep[][]{Moutard2016a}.
\textit{Ram-pressure stripping} \citep{Gunn1972}, generally put forth, would therefore require to be associated with \textit{tidal stripping harassment} \citep{Moore1996} to simultaneously shut star formation down and alter morphology or, alternatively, one may assign the quenching of low-mass galaxies to a major role of mergers within large-scale structures \citep[e.g.,][]{Schawinski2014}.
\end{itemize}

Still, the quenching mechanisms may be quite different depending on the scale of the structures involved in environmental quenching \citep[groups or clusters; e.g.,][]{Lin2014}. 
While our analysis confirmed the role of environment in the fast quenching of low-mass galaxies, the characterisation of the scale at which environmental quenching of low-mass galaxies operates would allow us to specify the physical mechanisms at play.

\section*{Acknowledgements}

We gratefully acknowledge the anonymous referee, whose comments helped in improving the clarity of the paper.
This research was supported by the ANR Spin(e) project (ANR-13-BS05-0005, \texttt{http://cosmicorigin.org}) and by a Discovery Grant from the Natural Sciences and Engineering Research Council (NSERC) of Canada.
This research makes use of the VIPERS-MLS database, operated at CeSAM/LAM, Marseille, France. This work is based in part on observations obtained with WIRCam, a joint project of CFHT, Taiwan, Korea, Canada and France. The CFHT is operated by the National Research Council (NRC) of Canada, the Institut National des Science de l'Univers of the Centre National de la Recherche Scientifique (CNRS) of France, and the University of Hawaii. This work is based in part on observations made with the Galaxy Evolution Explorer (GALEX). GALEX is a NASA Small Explorer, whose mission was developed in cooperation with the Centre National d'Etudes Spatiales (CNES) of France and the Korean Ministry of Science and Technology. GALEX is operated for NASA by the California Institute of Technology under NASA contract NAS5-98034. This work is based in part on data products produced at TERAPIX available at the Canadian Astronomy Data Centre as part of the Canada-France-Hawaii Telescope Legacy Survey, a collaborative project of NRC and CNRS. The TERAPIX team has performed the reduction of all the WIRCAM images and the preparation of the catalogues matched with the T0007 CFHTLS data release.




\bibliographystyle{mnras}
\bibliography{Moutard2017.bib} 







%


\bsp	
\label{lastpage}

\end{document}